\documentclass[11pt,a4paper]{article}
\usepackage{jheppub}

\usepackage[T1]{fontenc}
\usepackage[utf8]{inputenc}
\usepackage{latexsym}
\input{definitions.command}

\title{Anderson Localization in high temperature QCD: background configuration properties and Dirac eigenmodes}
\author[a]{Guido Cossu,}
\emailAdd{cossu@post.kek.jp}
\author[a,b]{Shoji Hashimoto}
\emailAdd{shoji.hashimoto@kek.jp}
\affiliation[a]{Theory Center, IPNS, High Energy Accelerator Research Organization (KEK),  Tsukuba, Ibaraki 305-0810, Japan}
\affiliation[b]{School of High Energy Accelerator Science, The Graduate University for Advanced Studies (Sokendai), Tsukuba 305-0810, Japan}

\abstract{
We investigate the properties of the background gauge field configurations that act as disorder for 
the Anderson localization mechanism in the Dirac spectrum of QCD at high temperatures.
We compute the eigenmodes of the M\"obius domain-wall fermion operator on configurations
generated for the $SU(3)$ gauge theory with two flavors of fermions, in the temperature range $[0.9,1.9]T_c$.
We identify the source of localization of the eigenmodes with gauge configurations that are self-dual 
and support negative fluctuations of the Polyakov loop $P_L$, in the high temperature sea of $P_L\sim 1$.
The dependence of these observations on the boundary conditions of the valence operator is studied. 
We also investigate the spatial overlap of the left-handed and right-handed projected eigenmodes in correlation 
with the localization and the corresponding eigenvalue.
We discuss an interpretation of the results in terms of monopole-instanton structures. 
}

\begin{document}
\maketitle

\section{Introduction}\label{sec:Introduction}

Quantum chromodynamics (QCD) is an inherently non-perturbative description of the strong force 
that governs the interaction of quarks. 
It has proven to be valid at zero temperature 
and to predict the phase transition to the quark-gluon plasma phase that occurs at high temperature. 
This phase transition separates the regime 
of confining and spontaneously broken chiral vacuum from the deconfining and chirally symmetric one.
In the case of fermions in the fundamental representation, the deconfinement temperature $T_d$ and 
the chiral restoration temperature $T_c$  are remarkably close to each other, suggesting a 
deeper connection between two otherwise unrelated properties of the vacuum.
There is still no widely accepted explanation for the correlation between the two transitions.\footnote{In the case 
of physical fermion masses for 2+1 flavors the two transitions
at the crossover are actually separated~\cite{Aoki:2006we}, but nevertheless they are very close.} 

The eigenvalues and eigenvectors of the QCD Dirac operator are important tools to probe the properties of the vacuum.
They control the low energy physics and chiral restoration.
A significant example of their relevance is the well-known Banks-Casher relation \cite{Banks:1979yr} 
that relates the spectral density $\rho(\lambda)$ at near-zero virtuality $\lambda \sim 0$
to the chiral condensate $\Sigma$, i.e. $\Sigma = \pi \rho(\lambda=0)$, in which the thermodynamical and massless limit are assumed. 
The spectral densities near the origin on the two sides of the chiral phase transition are radically different. Therefore, 
it is expected that the properties of the related eigenmodes reflect this change.
In this study we examine the relation between the lowest eigenmodes of the Dirac operator and their 
supporting background gauge configurations in order to shed some light on the possible
connection between confinement and chiral symmetry.

Another set of important characteristics of the Dirac spectrum are the local fluctuations of the eigenvalues in the bulk and the localisation 
of the corresponding eigenmodes.
The local fluctuations in the Dirac spectrum at high temperatures are quite different from their low temperature counterparts.
The high temperature Dirac spectrum experiences a transition between a low-energy region, where the eigenvalues are uncorrelated, and the 
higher part of the spectrum that exhibits non-trivial Random Matrix Theory (RMT) type of correlations 
(see \cite{PhysRevLett.105.192001, PhysRevD.86.114515, Giordano:2013taa} and Section~\ref{sec:DiracSpectrum}).
The critical energy at which the quantum transition occurs, $\lambda_c$, is called the mobility edge.  
The distribution of the low eigenvalues for the low temperature region is expected 
to agree with RMT models by the presence of a chiral condensate. 
The actual symmetry class in RMT depends on the Dirac operator discretization, 
the gauge group and the dimensionality of the space~\cite{Verbaarschot:2000dy}. 

There is a tight connection between the local spectral correlations and the localisation of the modes.
It has been proven using staggered and overlap fermions 
in~\cite{PhysRevLett.105.192001, PhysRevD.86.114515,Bruckmann:2011cc, Giordano:2013taa} 
(and we provide a confirmation using domain-wall fermions in Section~\ref{sec:DiracSpectrum})
that the lowest modes of the high temperature spectrum are localised below the threshold $\lambda_c$.
Above the critical eigenvalue the modes are delocalised, occupying the whole available volume and scaling accordingly. 
This behaviour is parallel to the Anderson localisation mechanism \cite{Anderson:1958vr} that explains the 
transition from a metallic phase (delocalised electron wave functions) to an insulator (localised modes) 
by the increasing density of impurities in a crystal. In the localised phase the wave functions are no more
extended waves (Bloch waves in the limit of zero disorder) but are described 
by an exponential decay with a typical scale called the localisation length.
Thus the QCD vacuum at high temperature is acting as an insulator for the lowest modes. 
The transition between the two regimes is known as Anderson transition 
and it is a quantum transition of second order for more than two dimensions~\cite{Abrahams:AndersonLocalization}. 
It is driven by the amount of impurities (disorder) in the crystal.

The authors of \cite{Giordano:2013taa} have shown that the scaling critical exponents for the QCD spectrum match the expectation from the
three dimensional Anderson Hamiltonian model, confirming that there is an Anderson-like mechanism in action in high temperature QCD.
The critical point depends on the temperature~\cite{Pittler:2014qea}, an indication that the disorder is depending on the temperature too. 
Moreover, the low temperature phase eigenmodes exhibit a metallic behaviour, 
being delocalised at all energies. There is no critical $\lambda_c$ in the spectrum in this regime.
The interesting conjecture is whether the opening of a localised region in the spectrum 
is connected to the chiral phase transition \cite{GarciaGarcia:2006gr, GarciaGarcia:2005vj, Giordano:2013taa, Pittler:2014qea, Giordano:2016cjs}.\footnote{
Anderson localisation has been considered in lattice QCD also in the context of construction of chiral fermions, domain-wall and overlap, regarding 
the localisation of the kernel eigenmodes and its effect on the chirality of the discretised action (see for example~\cite{Aoki:2001su, Golterman:2003qe}).}

The attractive question, once we assume Anderson localisation in QCD, is what are the impurities, or the disorder, in the 
background gauge configurations that trigger the localisation of the eigenmodes. 
How is the diffusion of fermions affected by the impurities? 
Are these related to the chiral transition and/or deconfinement? 
This is the central topic of this study. 
Concerning this question, in recent years \cite{Bruckmann:2011cc} it has been shown for $SU(2)$ pure gauge theories 
and argued using spin models \cite{Giordano:2015vla} that the Polyakov loop is acting as a source of localisation.
In Section~\ref{sec:Analysis} we show numerical evidences for several temperatures 
and masses that the lowest modes cluster where the real part of the Polyakov loop is negative and this correlation 
increases with the temperature. Higher modes, delocalised, show no preference regarding the value of Polyakov loop. 
Furthermore, we study the relation of the eigenvalues with the local action $F_{\mu\nu}F_{\mu\nu}(x)$, the local 
topology $F_{\mu\nu}\tilde F_{\mu\nu}(x)$ and the variation of the boundary conditions for the fermion field. 
A dependence on the boundary condition of the spectrum at high temperature has been analytically proven 
(see for example~\cite{Bilgici:2009tx}) and it is another useful way of probing the properties of 
the background configurations. 
In Section~\ref{sec:LeftRight} we examine the left and right projected components of the non-zero 
eigenmodes and their relation with the corresponding eigenvalues. 

In the last section 
we propose an interpretation of the results on the gauge-impurities based on 
Bogomolny-Prasad-Sommerfield (BPS) monopole-instanton objects~\cite{Prasad:1975kr, Bogomolny:1975de}. 
We argue that all the properties observed 
are explained by molecules of self-dual monopoles, sometimes called dyons in the literature. 


We then draw conclusions and perspectives for future work to understand 
how strong is the relation between the deconfinement and the chiral phase transition.

\section{Dirac spectrum and localisation properties}\label{sec:DiracSpectrum}

\subsection{Numerical setup}\label{sec:NumSetup}
We study the properties of the eigenmodes of the M\"obius domain-wall fermion operator \cite{Brower:2005qw,Brower:2012vk} on configurations 
around the phase transition temperature.
The gauge configurations are generated with two degenerate flavors of M\"obius domain-wall dynamical fermions and tree-level improved Symanzik gauge action.
Stout smearing \cite{Morningstar:2003gk} is applied to the gauge links inserted in the fermionic action. A detailed account on the choice of parameters is found in~\cite{Hashimoto:2014gta}.
We simulated several ensembles in the temperature range $[0.9,1.9]T_d$ where $T_d$ is the deconfinement temperature ($T_d \simeq T_c$), and several masses.
The deconfinement transition temperature is estimated to be $T_d = 175(5)$ MeV from the measurement of the average Polyakov loop. 
The range of bare quark masses is $[2,40]$ MeV depending on the ensemble. 
We have three type of lattice sizes $16^3\times 8$, $32^3\times 8$ and $32^3\times 12$ to control the dependence on the volume and 
lattice spacing. Table~\ref{tab:ensembles} lists the ensembles presented in this work. 

We calculated, using the implicitly restarted Lanczos algorithm~\cite{saad-eig}, 
the eigenmodes of the massive hermitian Dirac operator $H(m) = \gamma_5 D^{4d}_{DW}(m) \equiv \gamma_5 D$ 
(which also defines the notation for $D$ from now on). 
Here the four-dimensional effective Dirac operator of the five-dimensional domain-wall fermion
is constructed by combining it with the Pauli-Villars operator \cite{Brower:2012vk}.
We employed the IroIro++ codeset~\cite{Cossu:2013ola} to generate and analyse the configurations.  

For our study of the left and right projected modes we need the eigenmodes of the non-hermitian operator, that guarantees that the expectation values of 
$\gamma_5$ are zero for non-zero eigenmodes.
In order to obtain the eigenmodes of the non-hermitian operator we assume that $D$ satisfies the Ginsparg-Wilson relation. This assumption 
is checked numerically and it is satisfied with enough precision for our purposes. The depth of the fifth dimension $L_s$ is chosen such that this requirement is satisfied.  We use only ensembles with lattice spacing smaller than 0.09 fm to constrain the possible violations, which are discussed in detail in~\cite{Cossu:2015kfa}. 
Given the subspace of paired eigenmodes $\phi_1$ and $\phi_2$ for the hermitian $H=\gamma_5 D$ ($H\phi_{1,2}= \pm\lambda\phi_{1,2}$, $\lambda \in \mathbb{R}$) 
the corresponding eigenmodes of $D$ are $\psi_{+,-}= (\phi_1 \pm i\phi_2)/\sqrt{2}$ so that 
the local norm and the local chirality are just the sum of the corresponding $\phi_{1,2}$ matrix elements. 
These are the modes studied in this work.

\begin{table}[tbp]
  \begin{center}
    \begin{tabular}{ l l l | r}
      \hline
      $\beta$ & $m$ & $N_s^3\times N_t\,(\times L_s)$ & $T$ (MeV)\\\hline 
      4.18      &   0.009       &$16^3\times 32\,(\times 12)$ & 0\\ \hline
      4.10*     &   0.01        &$16^3\times 8\,(\times 12)$ & 216(1)\\
      4.10*     &   0.01        &$32^3\times 8\,(\times 12)$ & 216(1)\\
      4.18      &   0.01        &$16^3\times 8\,(\times 12)$ & 257(1)\\
      4.18      &   0.01        &$32^3\times 8\,(\times 12)$ & 257(1)\\
      4.30      &   0.01        &$32^3\times 8\,(\times 12)$ & 330(3)\\ 
      4.30      &   0.005       &$32^3\times 8\,(\times 12)$ & 330(3)\\ \hline
      4.18      &   0.01        &$32^3\times 12\,(\times 16)$ & 171(1)\\
      4.23      &   0.01        &$32^3\times 12\,(\times 16)$ & 191(1)\\
      4.23      &   0.005       &$32^3\times 12\,(\times 16)$ & 191(1)\\
      4.23      &   0.0025      &$32^3\times 12\,(\times 16)$ & 191(1)\\
      4.24      &   0.01        &$32^3\times 12\,(\times 16)$ & 195(1)\\
      4.24      &   0.0025      &$32^3\times 12\,(\times 16)$ & 195(1)\\
      4.30      &   0.01        &$32^3\times 12\,(\times 16)$ & 220(2)\\
      4.30      &   0.005       &$32^3\times 12\,(\times 16)$ & 220(2)\\
      \hline
    \end{tabular}
    \caption{Lattice ensembles used in this study. Bare quark mass $m$ and the lattice size $N_s^3\times N_t\,(\times L_s)$ are given in lattice units. The ensembles $\beta=4.10$ with $N_t=8$, shown by (*) have higher violation of chirality than the one required for Section~\ref{sec:LeftRight} (see discussion in the text). They are only used in Section~\ref{sec:Localization} where the effect of violations on the localisation is irrelevant.}
    \label{tab:ensembles}
  \end{center}
\end{table}

\subsection{Eigenmode localisation}\label{sec:Localization}

A basic quantity to understand the localisation properties is the participation ratio (PR) that is defined as 
\beq
PR_n = \Bigl (V \int |\psi_n(x)|^4 d^4x \Bigr)^{-1}, \qquad \int |\psi_n(x)|^2d^4x  = 1
\eeq
where the $\psi_n$ is the $n$-th mode of the Dirac operator, normalised to 1. 
$PR_n$ is constant ($=1$) in the purely metallic phase with a completely delocalised eigenfunction, $|\psi_n(x)|^2 = 1/V$. 
It is, on the other hand, proportional to $1/V$ if the mode is completely localised ($\delta$-function on some $\bar x$), 
which implies an insulator phase. 
With this definition the PR represents the fraction of the total volume occupied 
by the support of the eigenmode $\psi_n$. Therefore the scaling with the volume 
of the eigenmodes will distinguish the localised and delocalised states. 
At the Anderson transition $\lambda_c$
the eigenmodes manifest a multifractal behaviour 
that has been observed also in lattice QCD~\cite{Ujfalusi:2015nha}.
This regime is characterised by multiple scaling exponents depending on the scale. 

\begin{figure}[ht]
  \centering
  \includegraphics[width=.49\columnwidth]{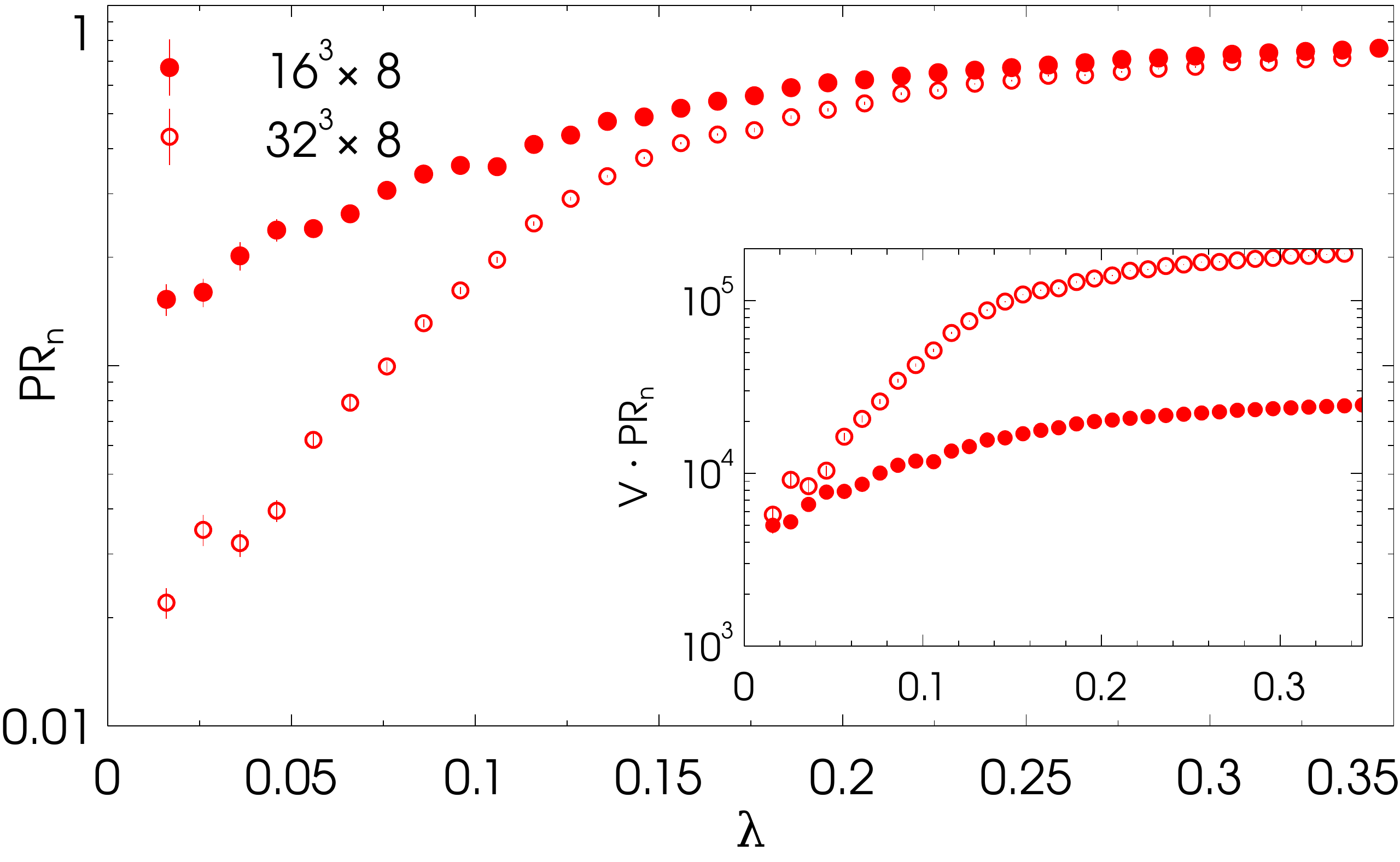}
  \includegraphics[width=.49\columnwidth]{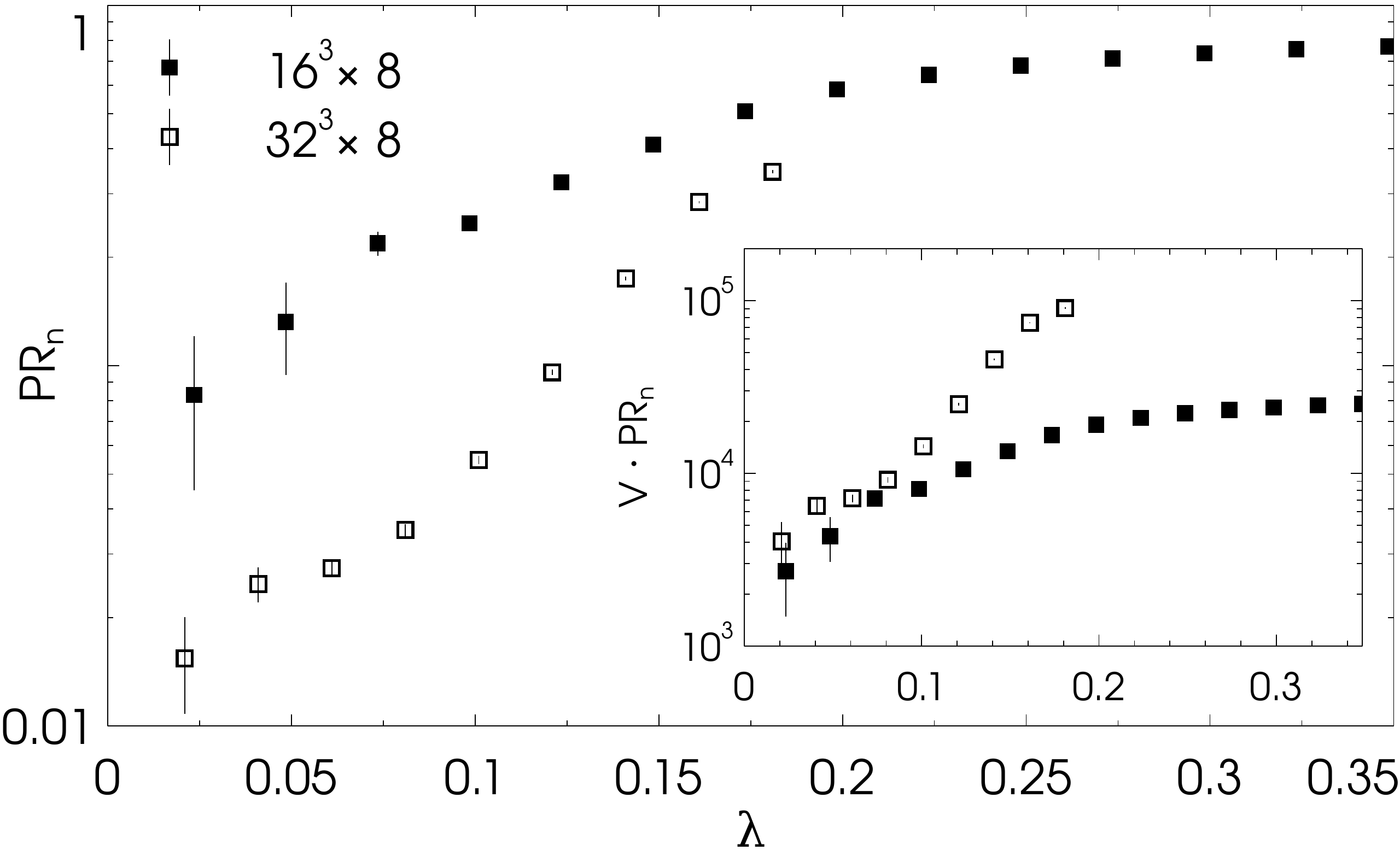}
  \caption{Scaling of the participation ratio for the ensembles $\beta=4.10$, $m=0.01$ (red circles) and $\beta=4.18$, $m=0.01$ (black squares), in log-scale.
Filled and empty symbols correspond to the $16^3\times 8$ and $32^3\times 8$ volumes.
The corresponding temperatures are $T=216, 257$ MeV and the bare quark mass is $17$ and $20$ MeV respectively. 
In the inset the participation ratio is multiplied by the volume.}
  \label{fig:IPRScaling}
\end{figure}

The scaling of low modes has been already studied in detail by Kovacs et al. 
using staggered fermions~\cite{PhysRevLett.105.192001, PhysRevD.86.114515}. 
We report here an example of our results as a warm-up and confirmation that 
the same phenomena observed in pure gauge ensembles 
and with different fermion discretisations~\cite{PhysRevLett.105.192001, PhysRevD.86.114515, Giordano:2013taa,Pittler:2014qea} 
are found also with chiral fermions on dynamical configurations. 
In Figure~\ref{fig:IPRScaling} we illustrate the scaling of the participation ratio for a couple of ensembles at $\beta=4.10, 4.18$
and $m=0.01$ where two different volumes $16^3\times 8$ and $32^3\times 8$ are available. 
It is evident in the high-mode region that the PR
approaches 1 and it is almost independent of the volume, a characteristic of delocalised modes. 
The scaling in the low-mode region is clearer in the inset, where the PR is multiplied by the total volume. 
The size of the lowest modes in this scale is independent of the volume implying 
a characteristic length: they are localised modes. 

\begin{figure}[ht]
  \centering
  \includegraphics[width=.49\columnwidth]{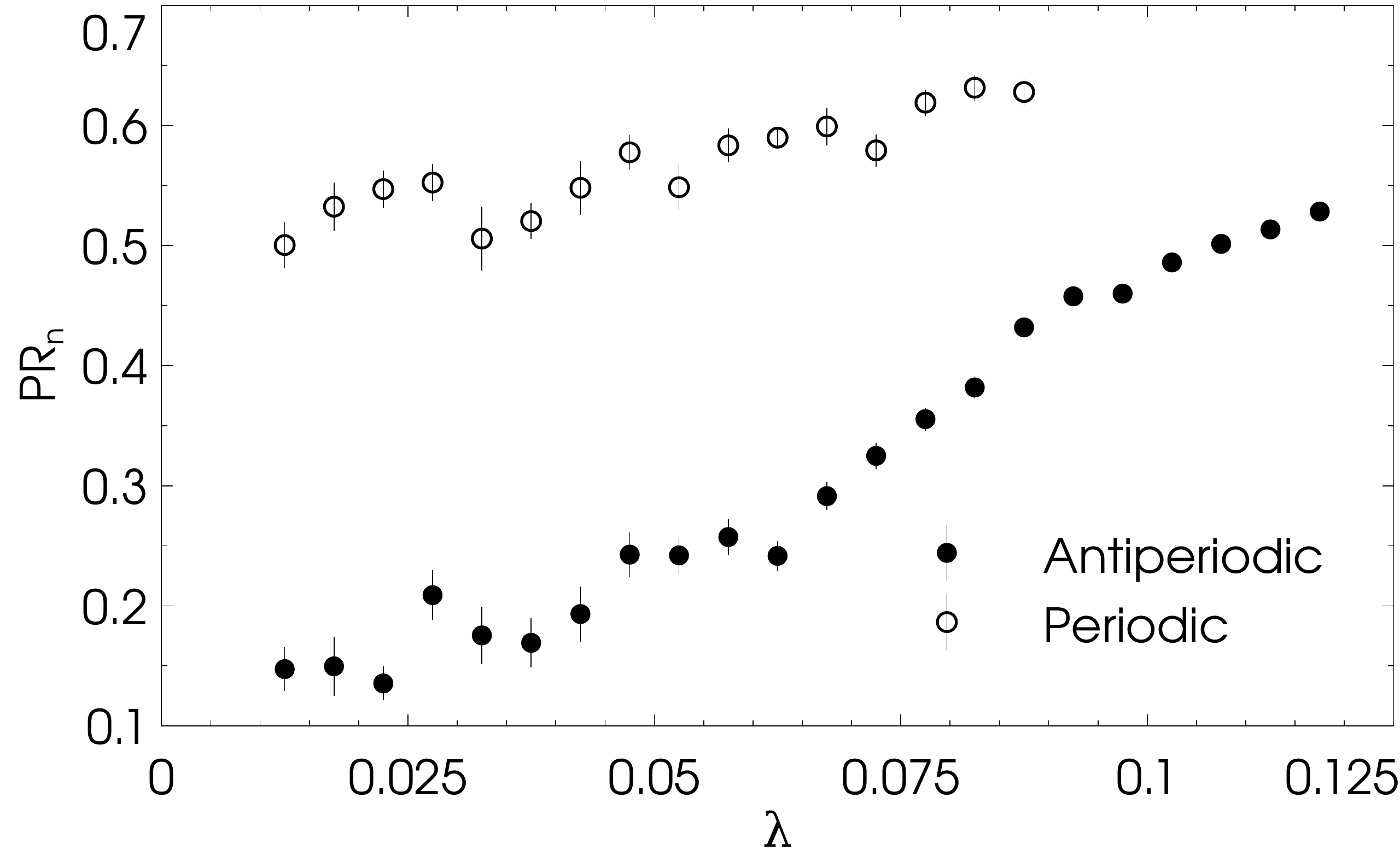}
  \caption{Dependence of the participation ratio on the boundary conditions for the ensemble $32^3\times 12$, $\beta=4.30$, $m=0.01$. 
The corresponding temperature is $T=220$ MeV and the bare quark mass is 26 MeV.}
  \label{fig:IPR_BC}
\end{figure}

The result for the dependence on the boundary conditions of the PR is reported in Figure~\ref{fig:IPR_BC}, for one ensemble. 
The localisation length of the lowest modes for the periodic boundary condition on the same configurations is quite different, 
occupying a larger fraction of the volume. It is an indication of a 
possible delocalisation but this conclusion cannot be inferred without a proper scaling analysis. 
In the following we will discuss other observables that complete the picture, 
showing delocalisation of the lowest modes when the boundary conditions are changed. 

\begin{figure}[tbh]
  \centering
  \includegraphics[width=.59\columnwidth]{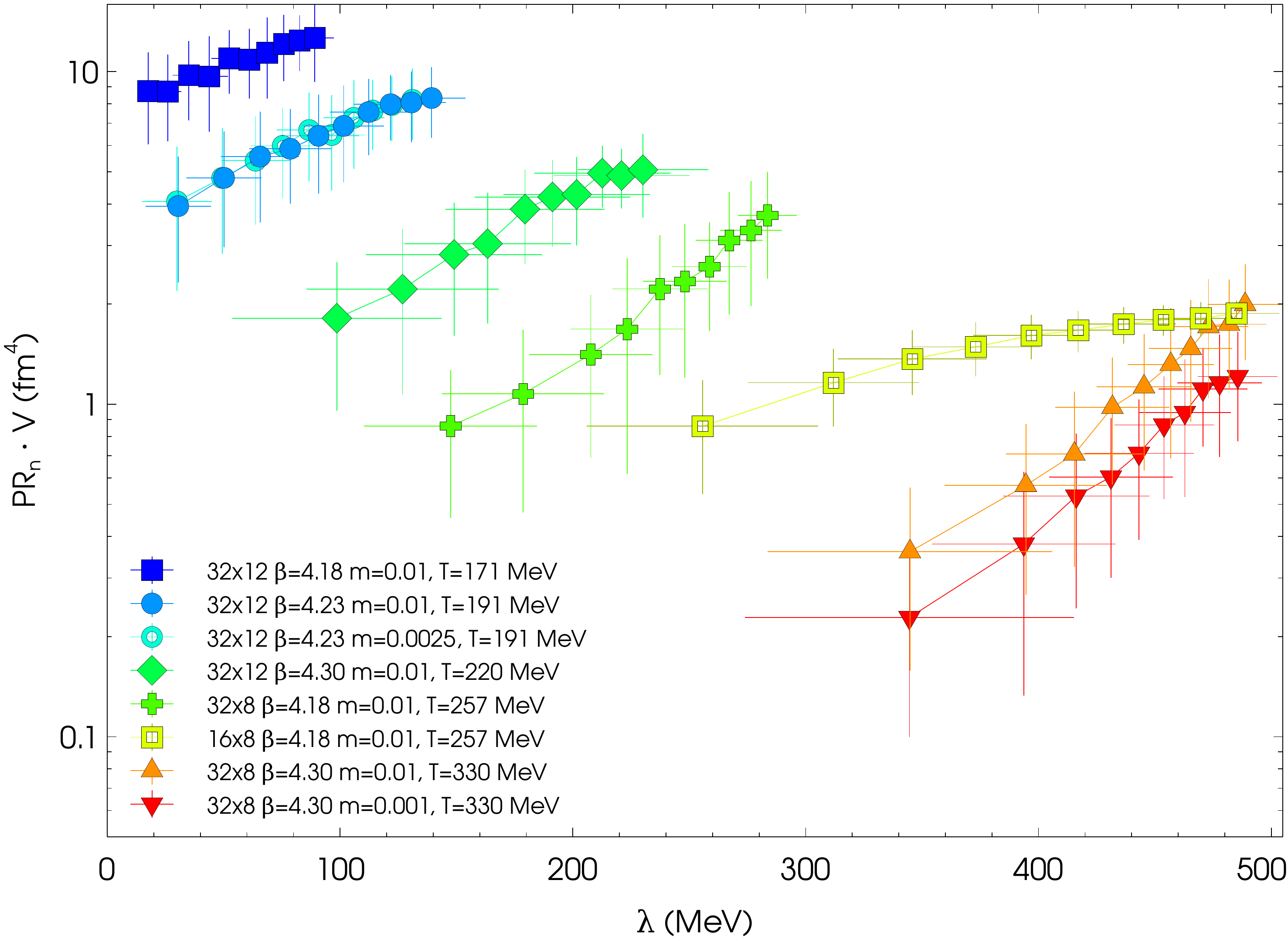}
  \caption{Dependence on the temperature of the physical size of 
the lowest 10 eigenmodes, averaged over the ensemble. 
The ensembles are ordered by increasing temperature. 
Volume effects visible in the higher modes for the $16^3\times 8$ ensemble (yellow box), 
that are limited by the total volume being $\sim 2.7$ fm$^4$. Compare to the correspoding $32^3\times 8$ ensemble (crosses).}
  \label{fig:PR_Lowmodes}
\end{figure}

\begin{figure}[hbt]
  \centering
  \includegraphics[width=.49\columnwidth]{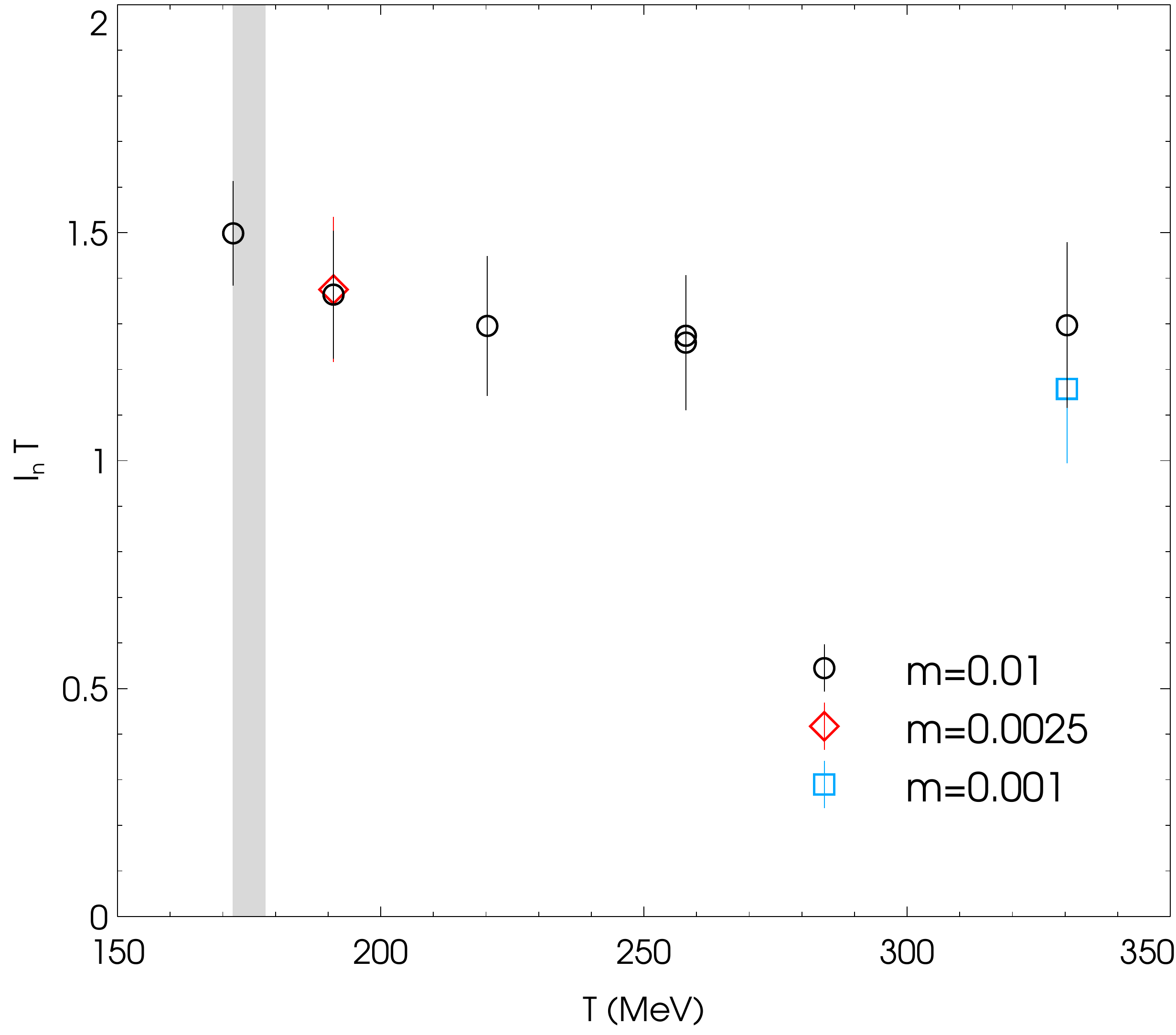}
  \caption{Temperature dependence of the lowest non-zero eigenmode localisation length $l_n = v_n^{1/4}$ times the temperature $T$.
The grey band indicates the estimated transition temperature, 175 MeV. The localisation length is governed by the temperature scale in a
 wide range of temperatures.}
  \label{fig:PR_FirstMode}
\end{figure}

The temperature dependence of the size of the localised modes is presented in Figure~\ref{fig:PR_Lowmodes}. 
Here we plot the physical size $v_n= PR_n\cdot V$ (in fm$^4$) of the first 10 non-zero eigenmodes against their eigenvalue, after averaging over the ensemble.  

The localisation length $l_n = v_n^{1/4}$ does not show an appreciable dependence on the mass of the valence quarks. 
The dimensionless quantity $l_nT$ is approximately constant $\sim 1.3$ for the lowest mode, see Figure~\ref{fig:PR_FirstMode}. 
This results indicates that the typical localisation length is governed by the temperature scale (see also~\cite{PhysRevD.86.114515} for an analogous result).

\subsection{Level spacing distribution}\label{sec:LevelSpacing}

The unfolded level spacing distribution (ULSD) is another trait that distinguishes 
the insulator and metallic phases in the Anderson model, together with localisation. 
The ULSD is the probability distribution $P(s)$ of 
\beq
s_i = \frac{\lambda_{i+1} - \lambda_i}{\langle \lambda_{i+1} - \lambda_i \rangle},
\eeq 
the distance of two consecutive eigenvalues in the unfolded spectrum of a matrix~\cite{RMTHandBook}. 
The spectral density is normalised to a constant, 1, to disentangle non-universal properties, a common procedure in this analyses.  
The ULSD can be analytically calculated for the three canonical RMT ensembles: 
Gaussian orthogonal ensemble (GOE), unitary (GUE), and symplectic (GSE). 
The eigenmodes in these ensembles are strongly correlated, with repulsion of nearest modes and suppression of distant modes. 
Completely decorrelated eigenvalues are scattered according to the Poisson distribution.
The insulator phase of the Anderson model, with strong disorder and localisation, has Poisson distributed eigenvalues
 while the weak disorder and metallic phase show delocalised modes following RMT. 
An interesting insight on this difference is given by the Bohigas-Giannoni-Smidth conjecture~\cite{PhysRevLett.52.1}.
It states that quantum systems with chaotic classical counterparts 
have a nearest-neighbour spacing distribution given by RMT whereas systems whose classical counterparts are integrable
obey a Poisson distribution (thus regular classical dynamics). 
Therefore a specific form of $P(s)$ is often taken as a criterion for the presence or absence of quantum chaos.

\begin{figure}[htb]
  \centering
  \includegraphics[width=.49\columnwidth]{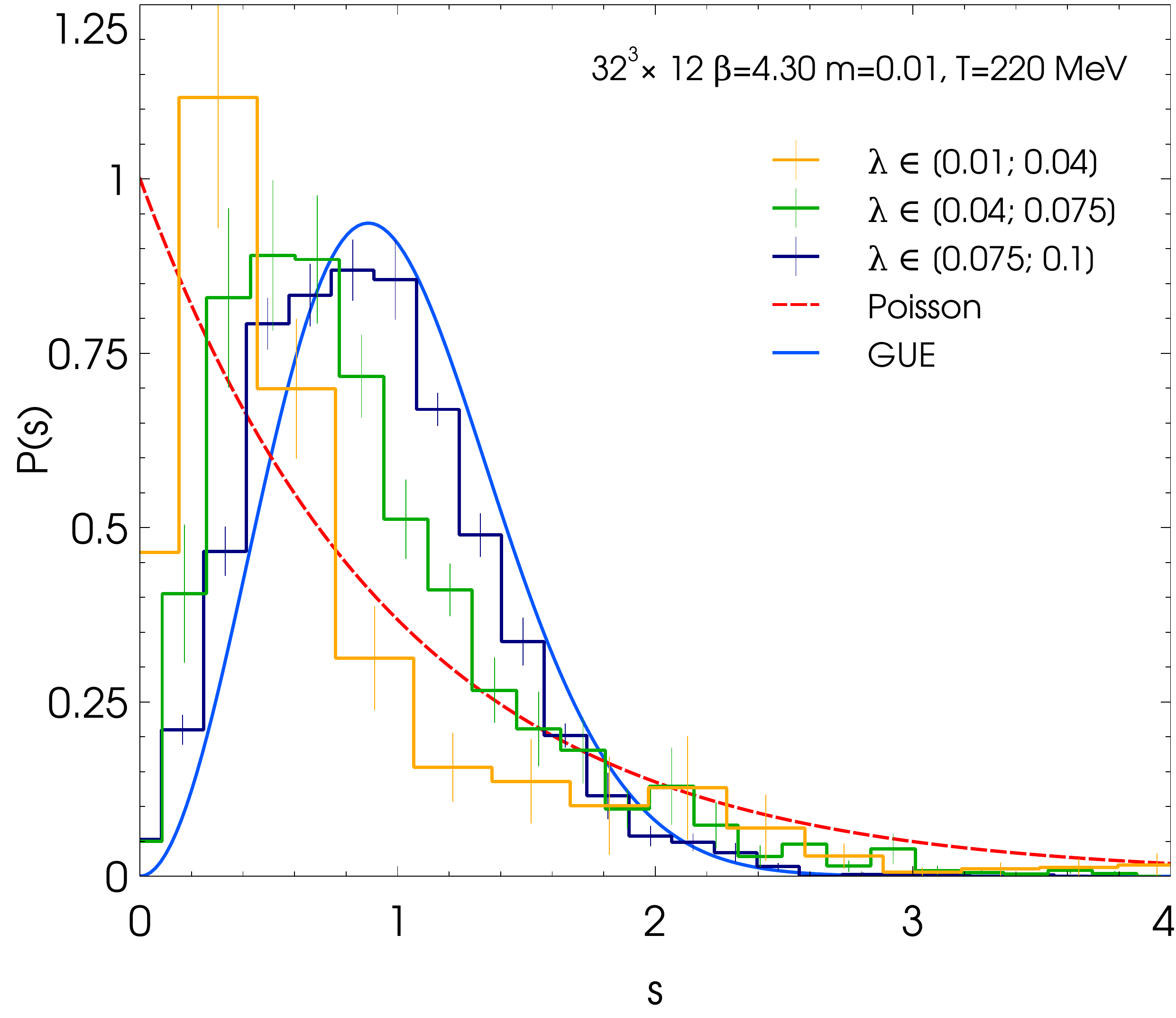}
  \includegraphics[width=.49\columnwidth]{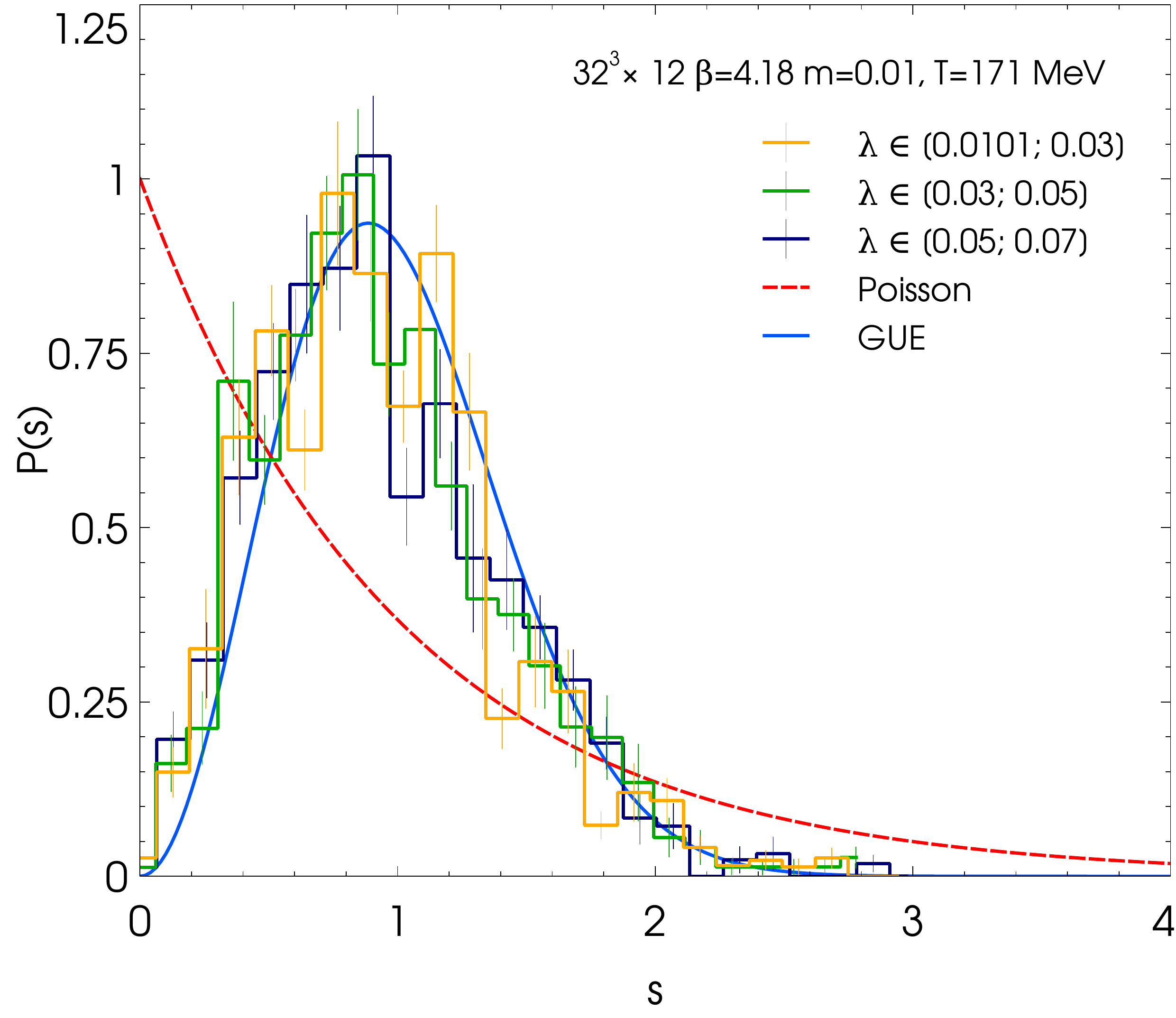}
  \caption{Distribution $P(s)$ of the level spacings for the unfolded Dirac spectrum in several sectors. The left and right panels are for the high temperature $T=220$ MeV and low temperature case $T=171$ MeV respectively. Continuous lines are the analytic predictions for totally uncorrelated eigenvalues (Poisson, red-dashed curve) and random matrix theory from the unitary ensemble (GUE, blue curve), with no free parameters.}
  \label{fig:Unfolding}
\end{figure}

By universality arguments the low temperature chirally broken phase in QCD is expected to follow the predictions of one of the RMT ensembles. 
For an $SU(3)$ chiral operator in the fundamental representation, 
the chiral unitary ensemble (chGUE) describes the low temperature level spacing.\footnote{GUE and chGUE are equivalent for the local fluctuations in the bulk. 
They differ for the $\epsilon$-regime predictions~\cite{Verbaarschot:2000dy}.} 
This has been extensively tested, for examples see the review~\cite{Verbaarschot:2000dy} 
and our data on the right panel of Figure~\ref{fig:Unfolding}. The $P(s)$ is in good accordance with the 
GUE distribution in three different regions of the spectrum, from the near-zero to the high eigenvalue region.
This is quite different from the distribution of the local fluctuations at high temperature, as we are going to show.

In the insulator phase the typical feature of the Anderson model is presence of the uncorrelated (Poisson) low modes. 
This is the kind of behaviour observed in the QCD spectrum for $T > T_c$ (see~\cite{PhysRevLett.105.192001, PhysRevD.86.114515, Giordano:2013taa,Giordano:2014qna} for previous detailed works). 
On the left panel of Figure~\ref{fig:Unfolding} we report one example from our ensembles, where $T \sim 1.2~T_c$. 
Three distributions are computed in different sectors of the spectrum, 
separated by the eigenvalues $\lambda=[0.01; 0.04], [0.04; 0.075], [0.075; 0.1]$. 
The near-zero sector follows the Poisson distribution (localised modes), while the bulk part is in accordance with RMT. 
In the middle an intermediate distribution is realised. Such kind of interpolating distributions were studied in~\cite{Nishigaki:1999zz}. 
These results confirm the idea that the system shows similarities with an Anderson model. 
This hypothesis has been quantitatively proved in~\cite{Giordano:2013taa} 
by a scaling analysis using staggered fermions. They showed that the critical
exponent for the correlation length $\nu$, is in agreement with the one of the 3d Anderson model.

\begin{figure}[htb]
  \centering
  \includegraphics[width=.6\columnwidth]{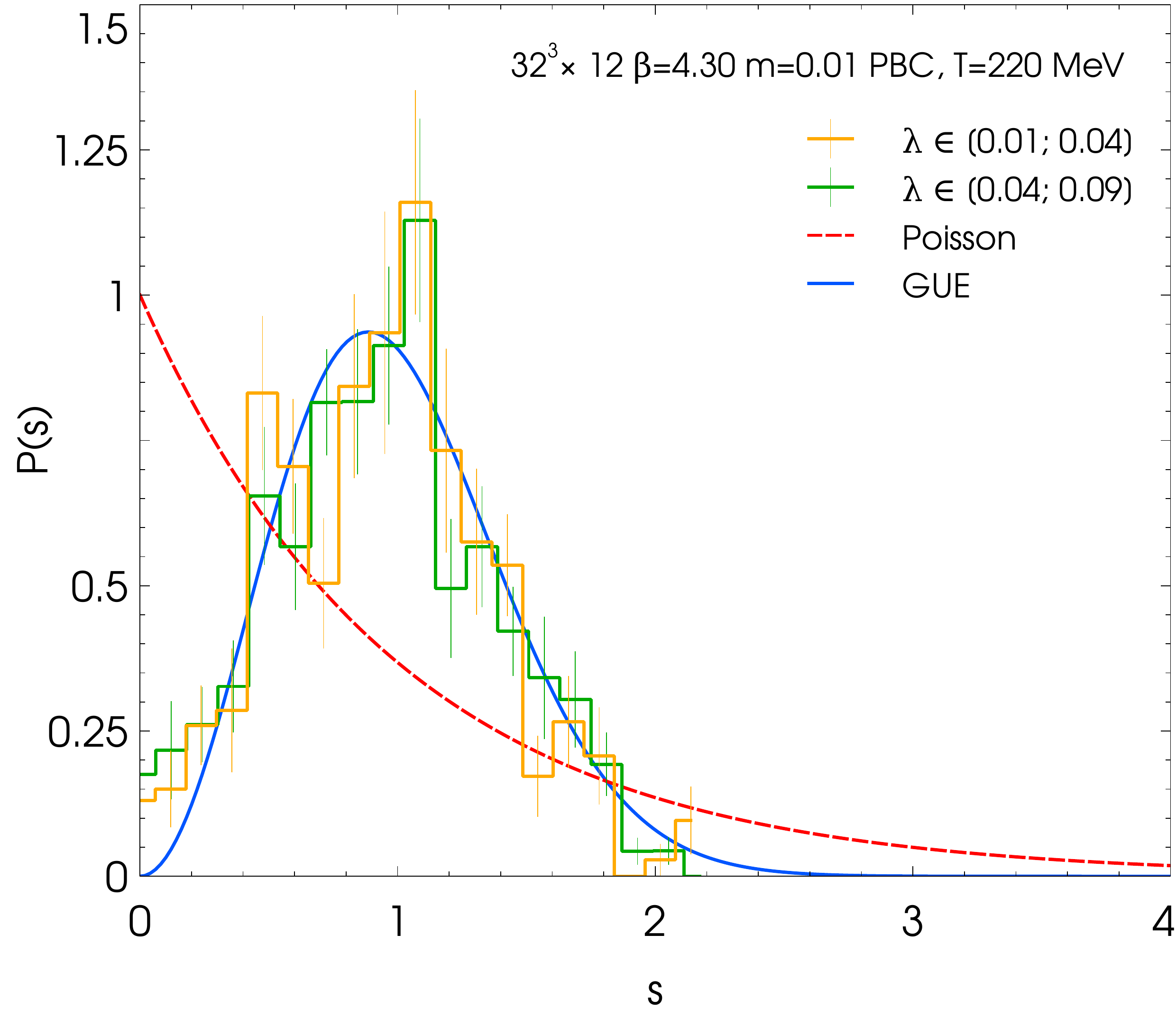}
  \caption{Distribution $P(s)$ of the level spacings for the unfolded spectrum at $T=220$ MeV and bare quark mass of 26 MeV with periodic boundary condition.  Continuous lines are as in Figure~\ref{fig:Unfolding}.}
  \label{fig:Unfolding_PBC}
\end{figure}

By changing the boundary condition on the same ensemble, Figure~\ref{fig:Unfolding_PBC}, 
the ULSD changes radically and it is in accordance with RMT, similar to the low temperature case. 
This is expected from our discussion on the PR. 
In our perspective it poses the question of how the eigenmode localisation depends on the boundary conditions. 
The background gauge field is unchanged, the lattice physical dimensions are unchanged, 
in particular we still have a compact dimension, but the modes are now delocalised. 
In the following we investigate this difference. 
Notice also that the spectral density has a gap in the anti-periodic case and
the gap disappears in the periodic case with non zero density at the origin. 
This is a clear signal of a chirally broken phase. 
A monopole instanton picture that addresses this behaviour is discussed in Section~\ref{sec:BPSMonopoles}.

\section{Numerical analysis of the background gauge field}\label{sec:Analysis}

We now discuss the numerical results on the background gauge field supporting the eigenmodes, 
with the intent of identifying the source of localisation. 
We address the interpretation later in Section~\ref{sec:BPSMonopoles}.
We are looking for a gauge invariant characterization of the background gauge field.

\subsection{Correlation with the local Polyakov line}\label{sec:PLBackgroundGauge}
We begin with the investigation of the correlation of the eigenmodes with the trace of the Polyakov line, or holonomy,
in a gauge theory with $N$ colors
\beq
P(x) = \frac{1}{N} \Tr L(x) = \frac{1}{N} \Tr \exp \Bigl (i \oint A_4 {\rm d}\tau \Bigr ) . 
\eeq
$\tau$ being the compact temporal dimension. 
This is a gauge invariant quantity related to confinement.
Similar studies on this observable were conducted 
by other groups on pure gauge configurations or using effective models~\cite{Bruckmann:2011cc, Giordano:2015vla}.

\begin{figure}
  \centering
  \includegraphics[width=.49\textwidth]{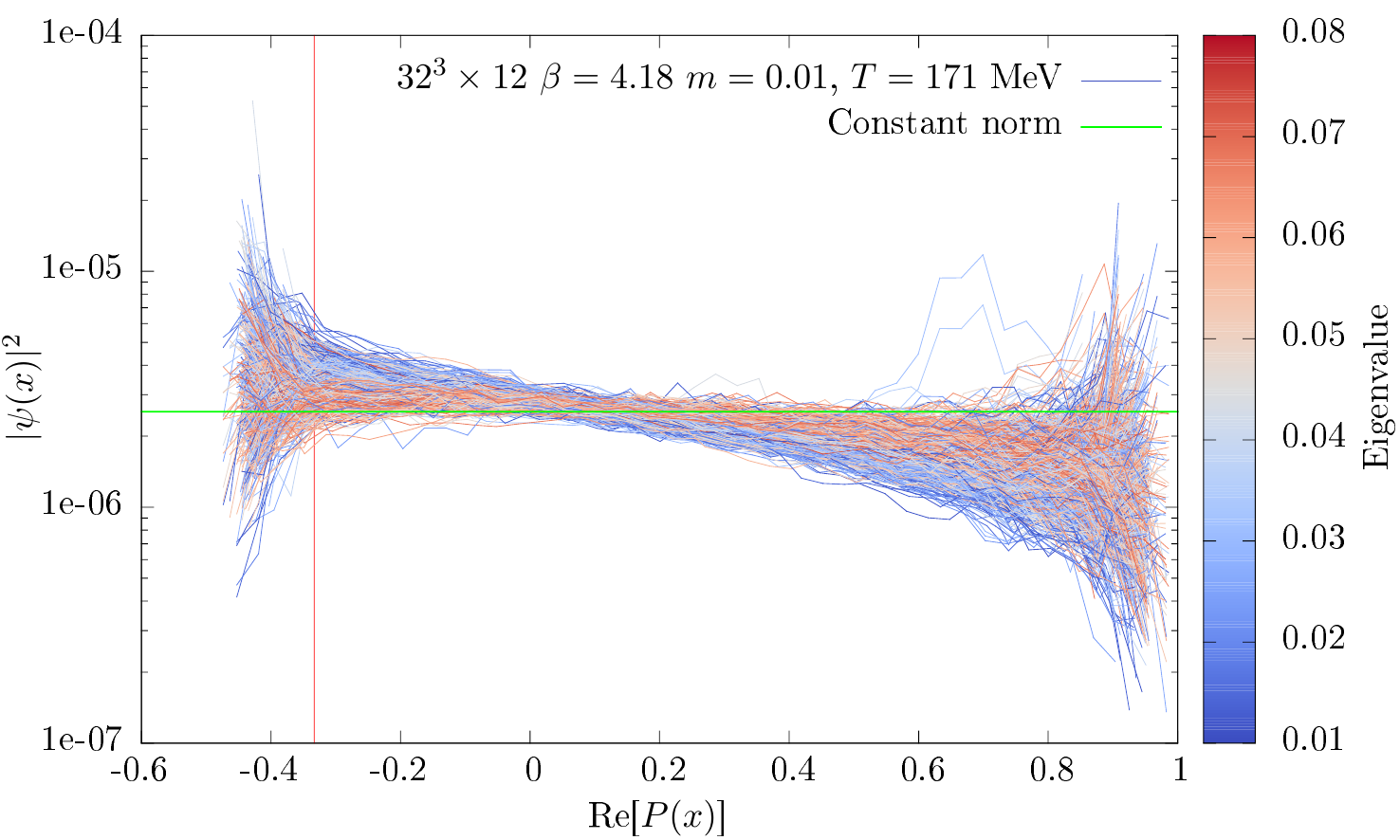}
  \includegraphics[width=.49\textwidth]{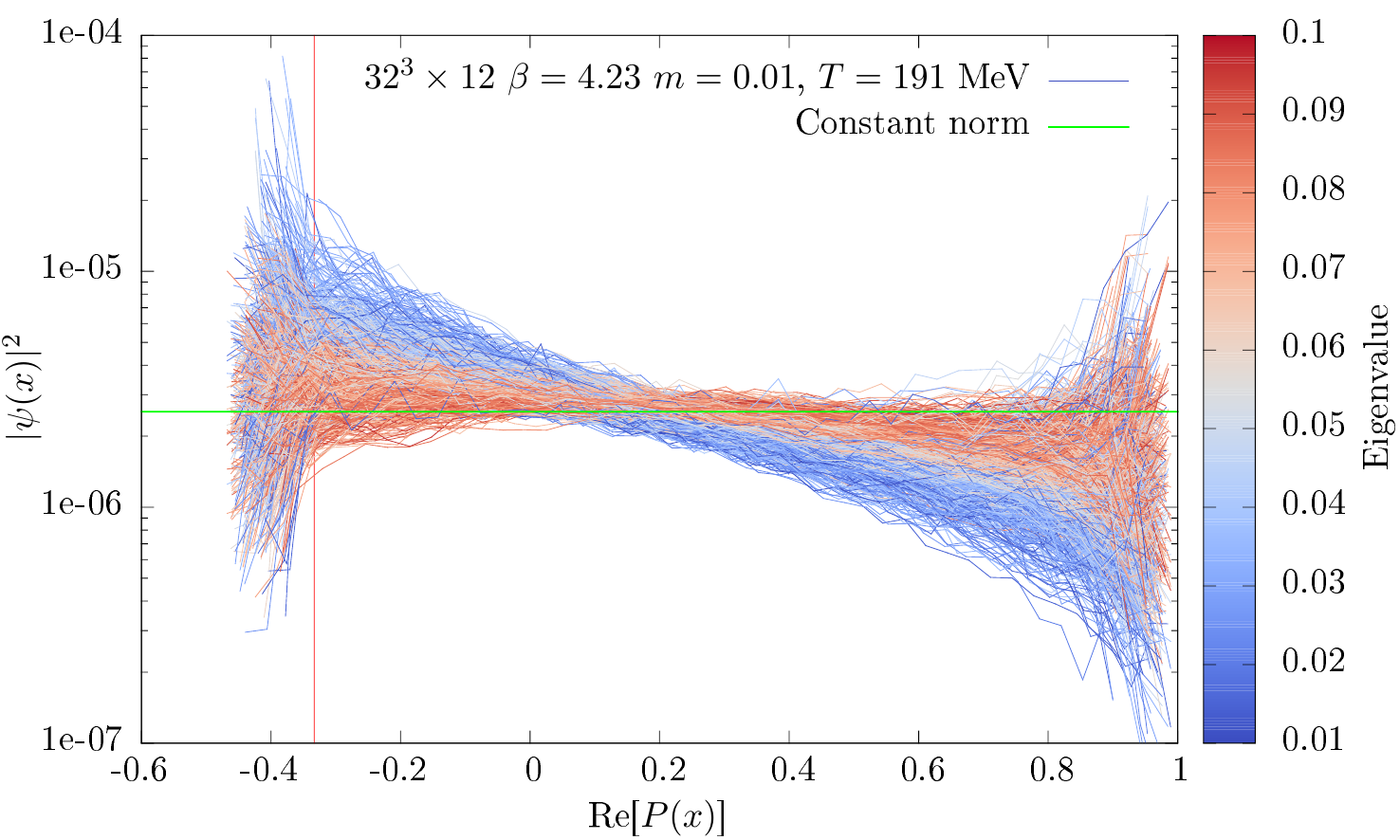}\\
  \includegraphics[width=.49\textwidth]{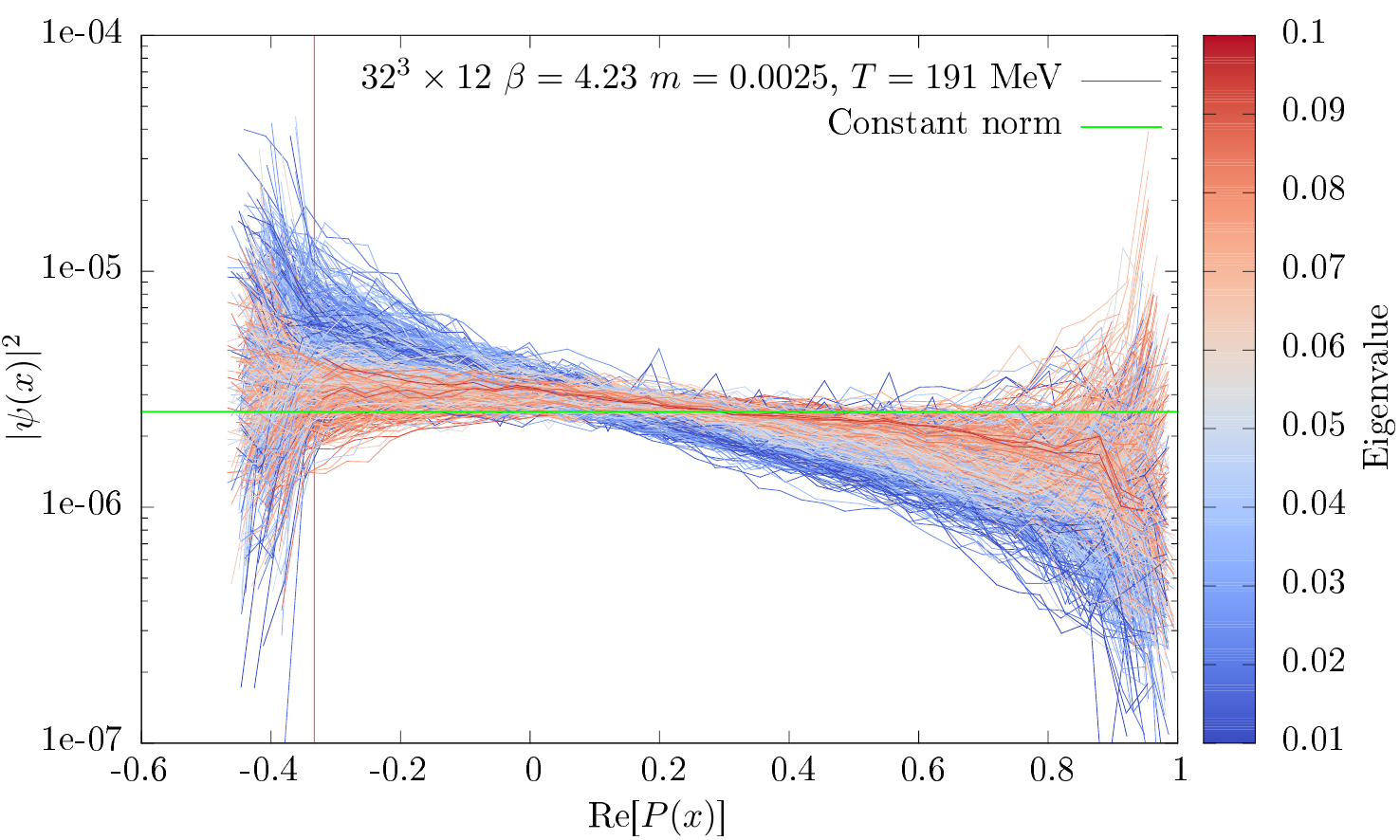}
  \includegraphics[width=.49\textwidth]{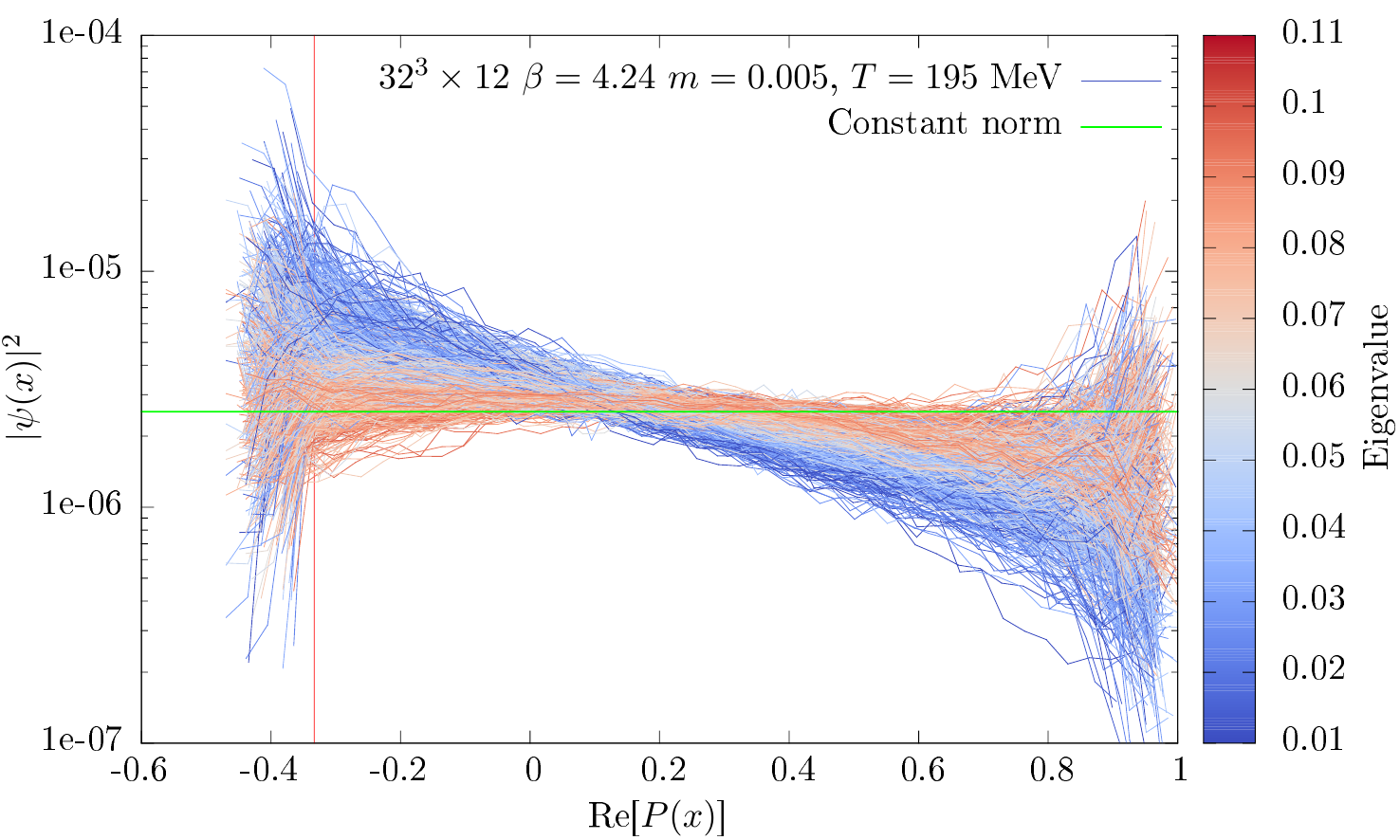}\\
  \includegraphics[width=.49\textwidth]{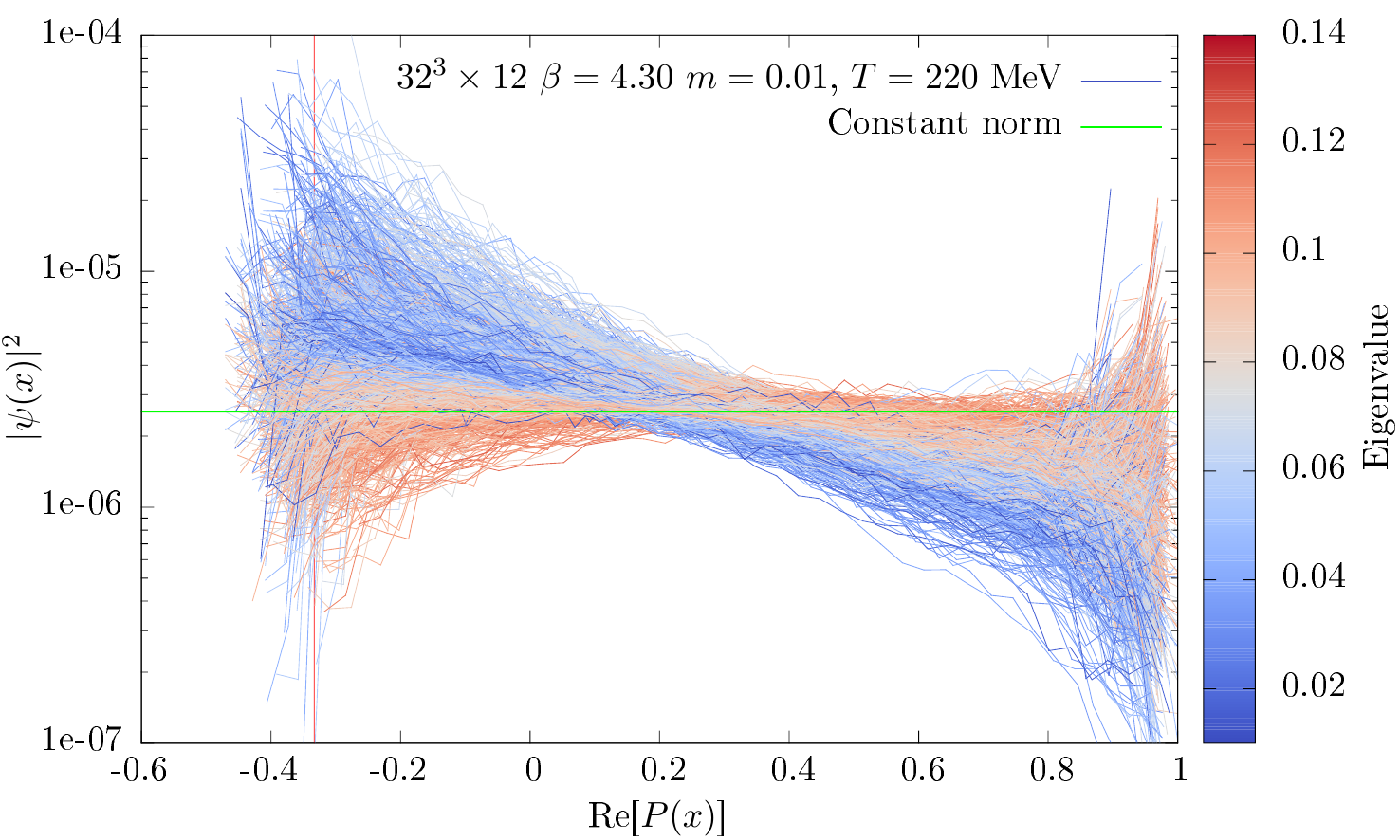}
  \includegraphics[width=.49\textwidth]{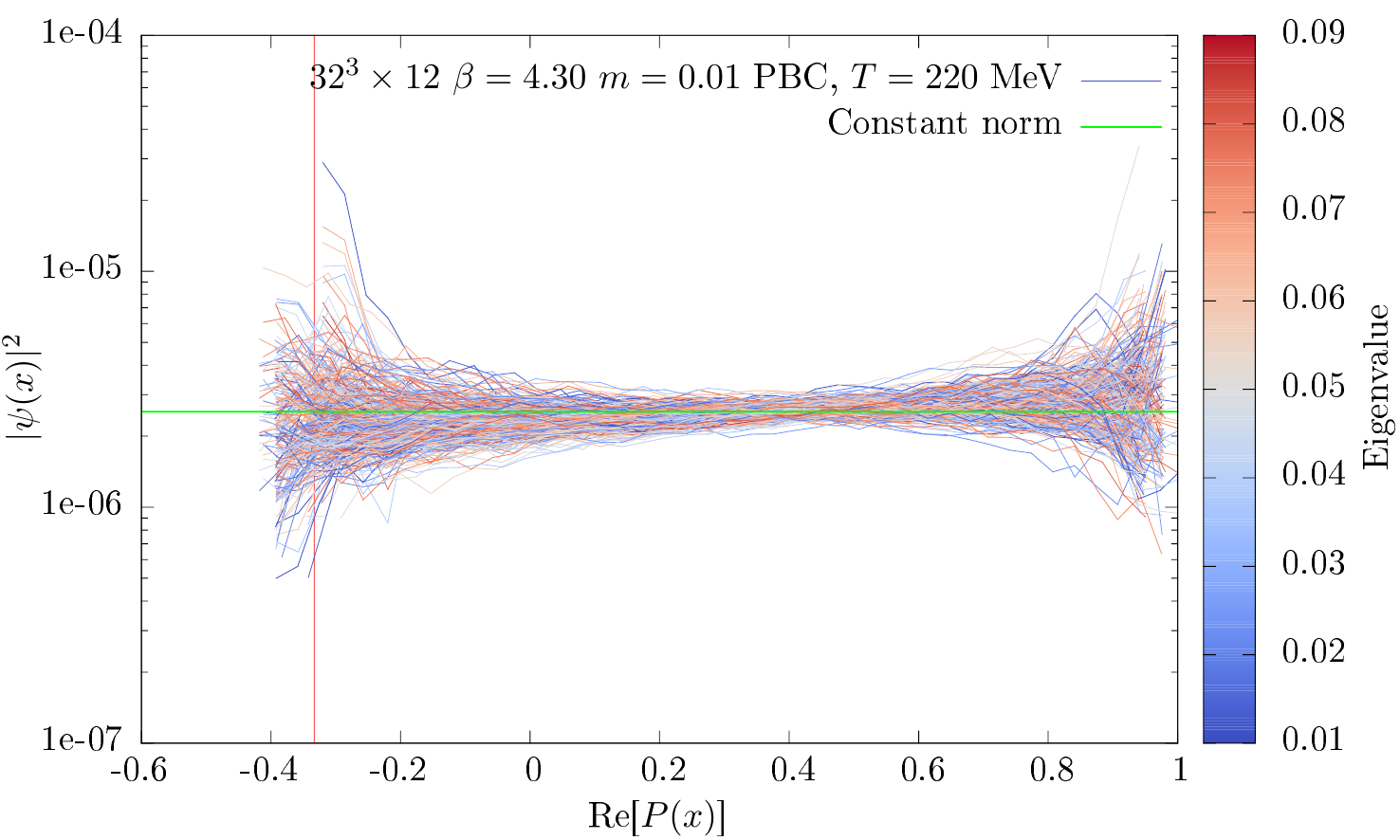}
  \caption{The local squared norm of the eigenmode in correlation with the local Polyakov line, real part. 
Temperature increases from left to right, top to bottom. The last panel is for eigenmodes with periodic boundary conditions. 
The colours represent the eigenvalue, as indicated by the colour bar attached to each plot. The green horizontal line represents a vector with a constant norm. Log scale for the norm axis is used. 
The vertical red line indicates the location of ${\rm Re} P = -1/3$.  }
  \label{fig:PLcorr}
\end{figure}

The Polyakov line is measured after suppressing UV fluctuations with the Wilson flow 
(other smoothing methods like cooling are equivalent~\cite{Bonati:2014tqa}). 
In the high temperature phase the spatial average of the Polyakov line $P(x)$ gives a finite value $\rightarrow 1$,
but we are interested in its local structure.
We then correlate, for several temperatures and masses, the local norm of the non-zero eigenmodes $|\psi(x)|^2$, 
with the real part of the Polyakov line trace. 
The results are reported in Figure~\ref{fig:PLcorr}. Every single line represents one eigenmode on one configuration. 
We average the norm of the mode in small intervals of the Polyakov line axis to clear up the presentation.
The colour scale represents the position in the spectrum of the eigenmode, see the colour bar. 
The green horizontal line stands for a normalised flat mode, everywhere constant. 
A mode with no correlation with the Polyakov line would follow this line. 

The plots show that all the modes just below the phase transition (upper left panel, $T=171$ MeV) 
have a mild dependence of $|\psi(x)|^2$ on the Polyakov line, notice the log-scale. 
This is in marked contrast with the highest temperature case (lower left panel, $T=220$ MeV).
At this temperature the lowest eigenmodes are localised, i.e. the norm is higher, where the Polyakov line is mostly negative. 
The highest modes, delocalised, show again no particular dependence on ${\rm Re}P(x)$. The norm of the low modes on the 
regions where ${\rm Re}P(x) \rightarrow -1/3$ increases with the temperature, see the sequence of the first five panels. 
The modes are more and more localised in these areas. 
These are the regions where two of the eigenvalues of the Polyakov loop are close to $-1$. In the majority of the volume 
at this temperature, on the other hand, the Polyakov loop eigenmodes are all similar and approach 1. 

We also consider the mass dependence of the correlation 
but we cannot observe any strong dependence in a range of masses spanning one order of magnitude,
although they are not shown here.

The last panel shows that by changing the boundary condition the correlations are washed out and all the modes are flattened.
This is another signal of the delocalisation associated to the periodic boundary condition.

These results indicate that all the localised modes are centred in the regions 
where the holonomy is far from its average for anti-periodic boundary condition. 
They repel the region where the Polyakov line is 1, which at high temperature is the most common value. 
The change of boundary condition to periodic cancels this localisation, even if the ``wrong'' $P$ islands are still present in the background. 
The modes are now delocalised and do not show any preference on the Polyakov line value, a situation similar to the low temperature case. 
The boundary condition affect the way the eigenmodes can diffuse in the volume. 
These observations will be useful in the discussion of Section~\ref{sec:BPSMonopoles}.

\subsection{Correlation with action and topology density}\label{sec:SQBackgroundGauge}
Two other relevant gauge invariant quantities that characterise the background configuration are 
the action density $s(x)=F_{\mu\nu}F_{\mu\nu}(x)$ and the topological charge density $q(x)= F_{\mu\nu}\tilde F_{\mu\nu}(x)$, 
without the canonical constant $8/32\pi^2$ in front. For a self-dual field $\tilde F = F$, the ratio $|q|/s = 1$. 
As with the Polyakov line, these variables are measured after smearing with the Wilson flow. 
The topological charge has been renormalised multiplicatively using a global fit on the whole ensemble~\cite{DelDebbio:2002xa}. 

\begin{figure}
  \centering
  \includegraphics[width=.49\textwidth]{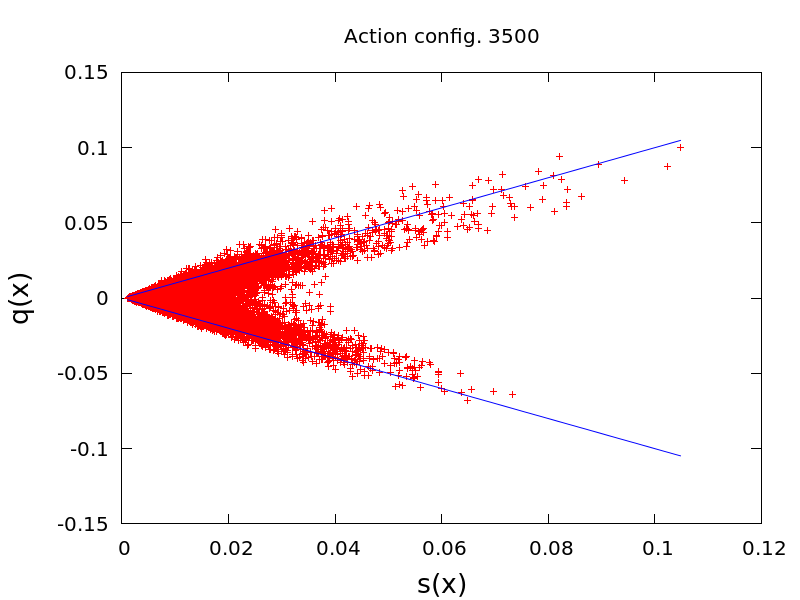}
  \includegraphics[width=.49\textwidth]{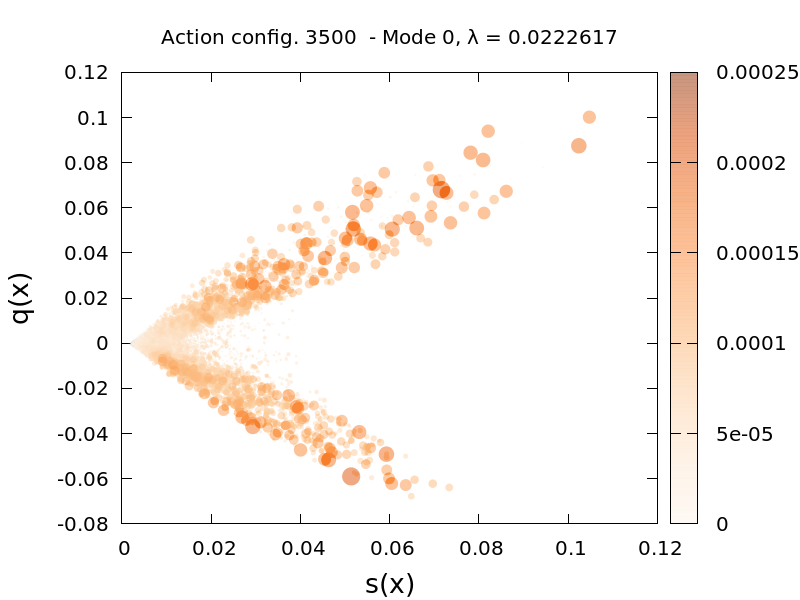}\\
  \includegraphics[width=.49\textwidth]{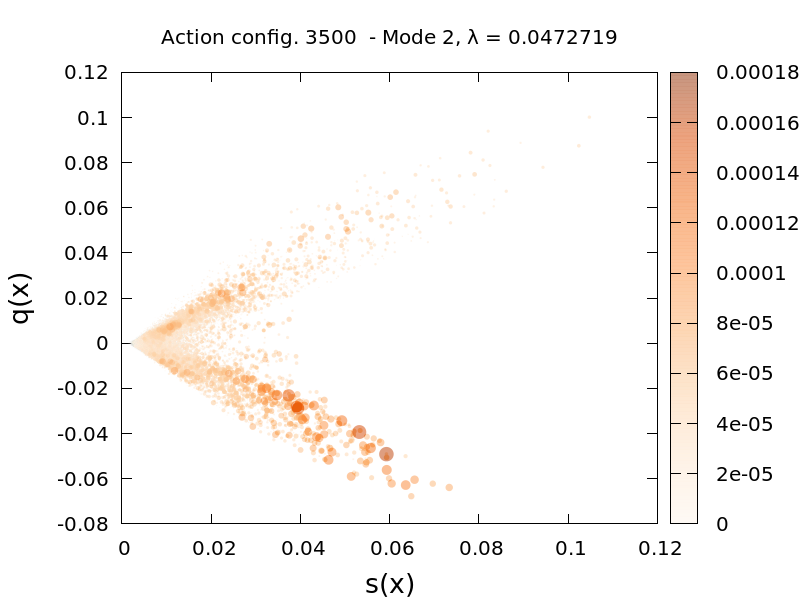}
  \includegraphics[width=.49\textwidth]{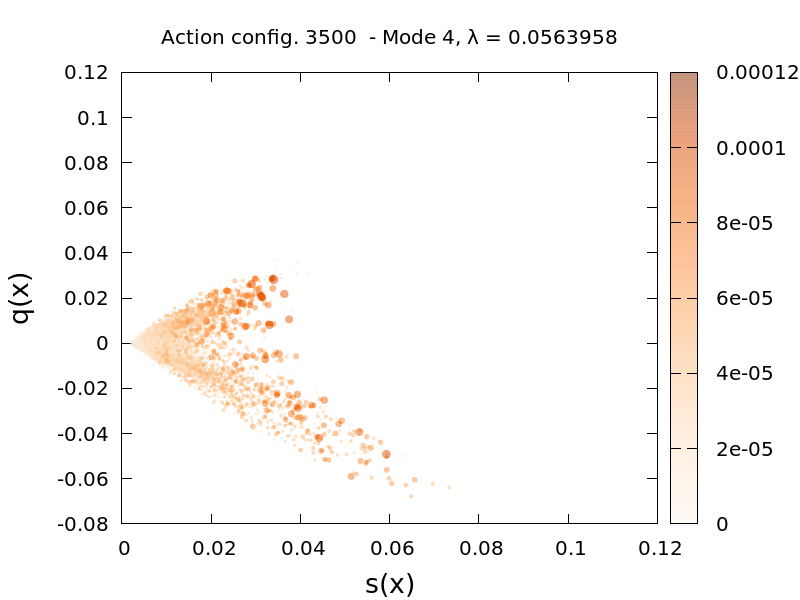}\\
  \includegraphics[width=.49\textwidth]{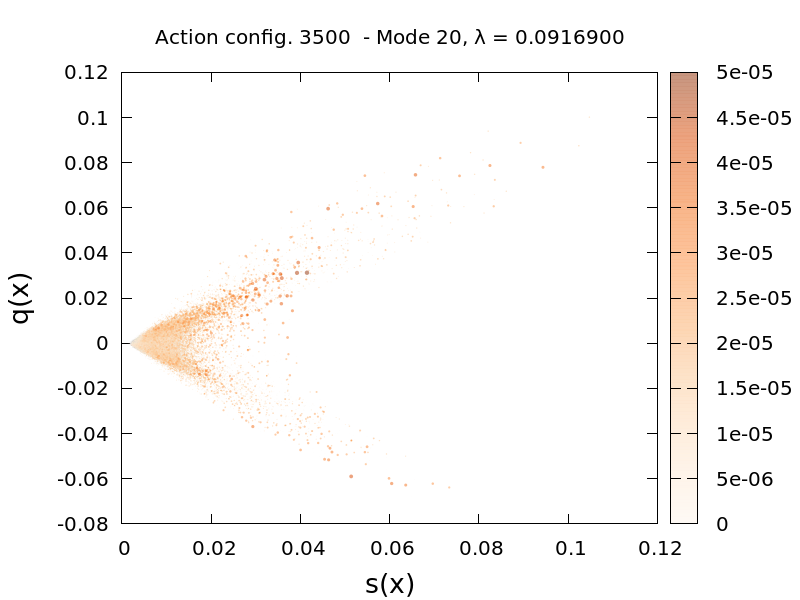}
  \includegraphics[width=.49\textwidth]{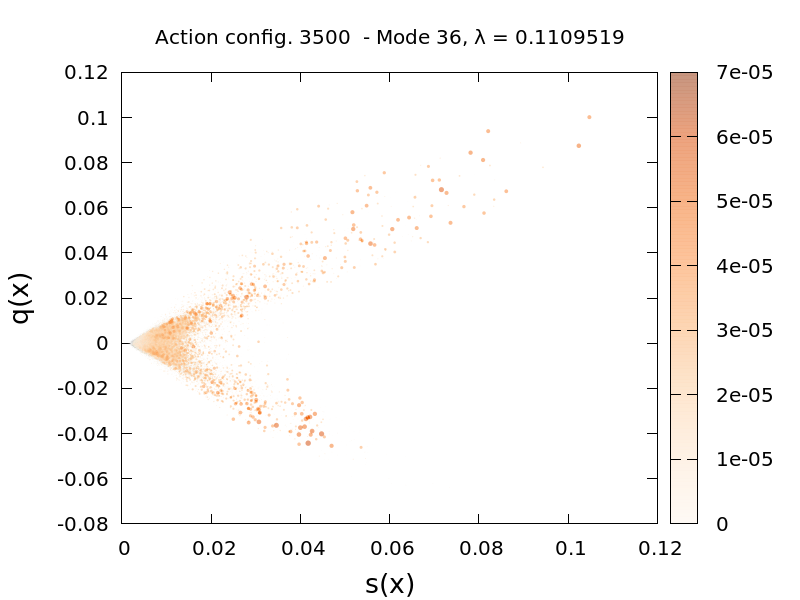}\\
  \caption{Correlations between the action density $s(x)$ and topological charge density $q(x)$ (first panel) and the eigenmode local norm, following panels. Configuration
number 3500 for the $32^3\times 12$, $\beta=4.30$, $m=0.01$ ensemble. This configuration does not contain zero modes.}
  \label{fig:Action}
\end{figure}

\begin{figure}
  \centering
  \includegraphics[width=.49\textwidth]{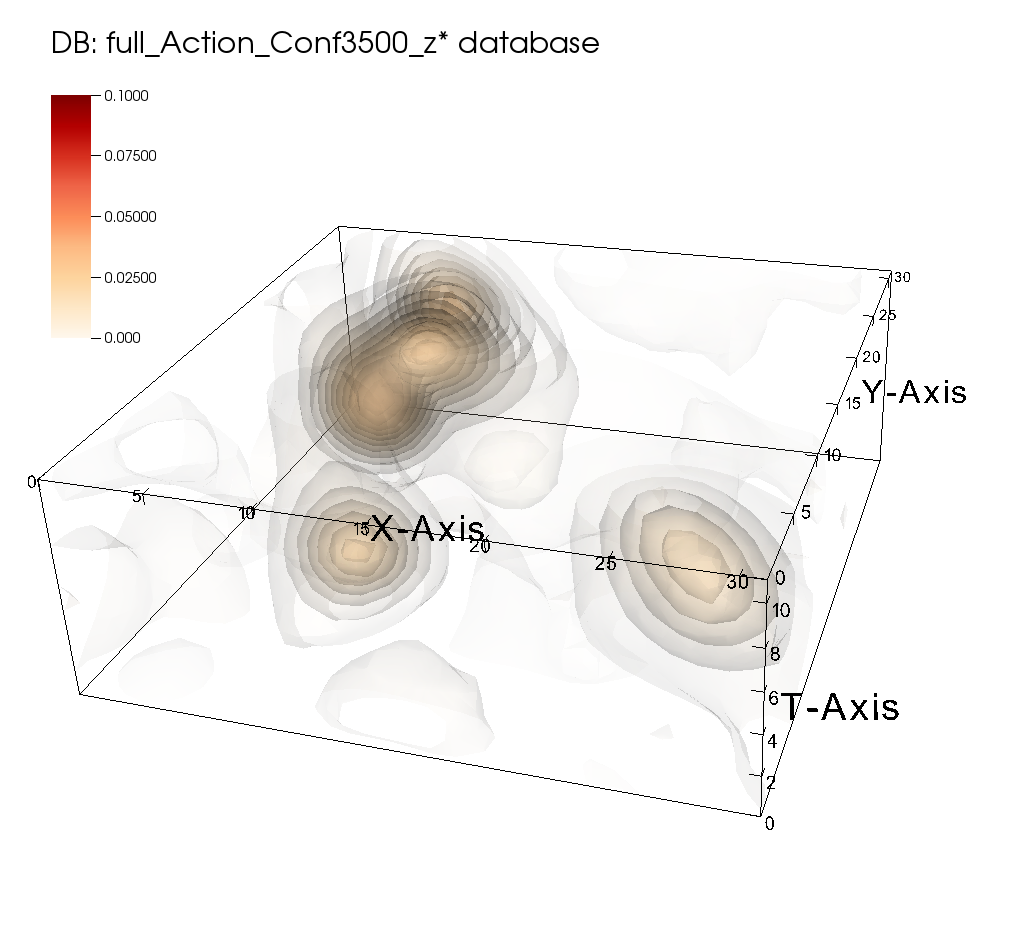}
  \includegraphics[width=.49\textwidth]{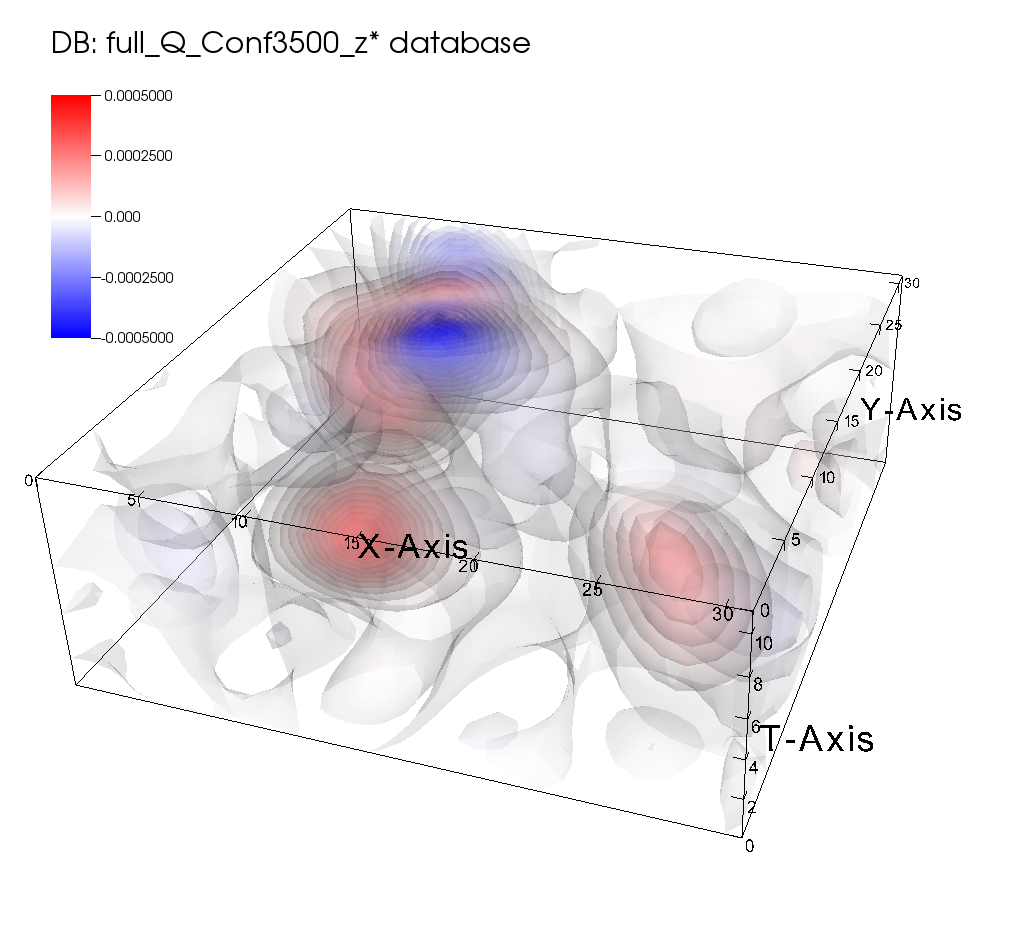}
  \caption{A three-dimensional section ($z=9$) of the action, left, and the topological charge, right, for the configuration n. 3500 of the ensemble $32^3\times 12$, $\beta=4.30$, $m=0.01$, same as Figure~\ref{fig:Action}.
The action is clustered in tube-like regions along the temporal direction, T-axis, with strong correlation among the time-slices. }
  \label{fig:Action3D}
\end{figure}

Before analysing the ensemble averages, it is instructive to look at one single configuration. 
In Figure~\ref{fig:Action} we present a typical situation for the eigenmode localisation in relation
to $s(x)$ and $q(x)$. 
Here we pick one configuration at high temperature ($T\sim 1.2~T_c$) and without zero modes. 
The first panel is a scatter plot of $q(x)$ versus $s(x)$ where every point corresponds to one lattice site.
It shows two branches that demonstrate that the regions where the action is larger coincide with the regions 
where the topological density $q(x)$ increases in absolute value. 
The branches follow the two continuous lines that are the sector secants, $s = |q|$. 
On the branches the action and the topological charge are the same on average: they represent almost self-dual fluctuations of the gauge field.
This is a common result on every configuration we analysed. 
Notice that several self dual fluctuations can be present in the configuration, though the plot does not capture this information.
Figure~\ref{fig:Action3D} depicts the actual three-dimensional distribution of action and topological charge in a z-slice where the action has the highest peak. 
Several peaks are visible besides the largest one. The action tends to cluster in a tube in the temporal direction. 
 
The following panels in Figure~\ref{fig:Action} add to the scatter plot the information of the eigenmode norm $|\psi_n(x)|^2$ on site $x$. The colour and the size of the 
dots represent the local norm of the eigenmode. Note that the top end of the colour range changes in the panels. 
The mode number increases from left to right, top to bottom. The lowest modes are localised on the branches, i.e. on the regions 
where the action and the topological charge are higher.  

\begin{figure}[htp]
  \centering
  \includegraphics[width=.7\columnwidth]{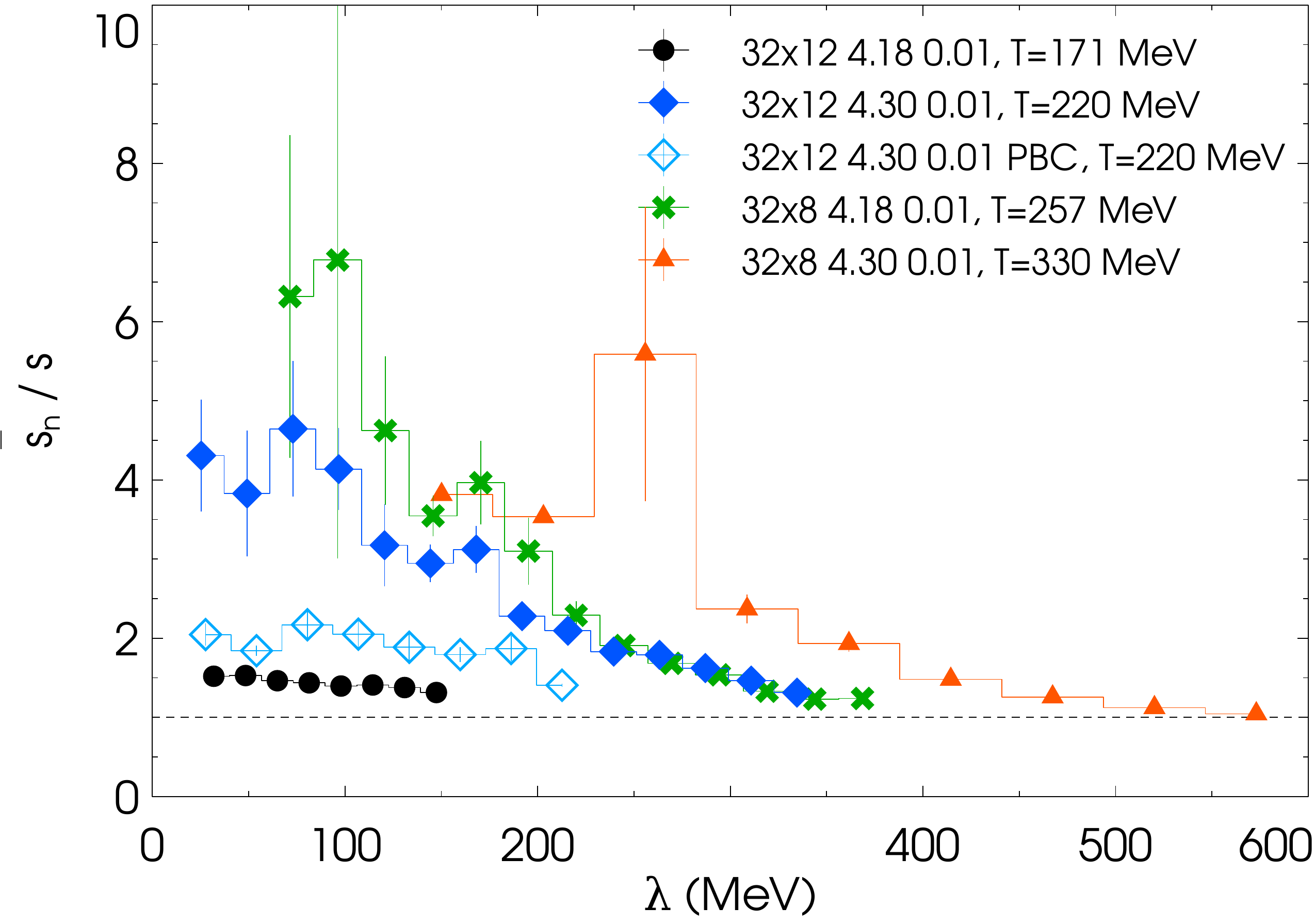}
  \caption{Ratio of the eigenmode weighted average of the action $\bar s_n$ and the average $s$, in correlation with the eigenvalue. 
Lowest modes favour higher action regions. Temperatures ranging from $[0.9, 1.9]~T_c$.}
  \label{fig:ActionDensityNorm}
\end{figure}

To systematically study the correlation with the action density of the eigenmodes we computed the following weighted average:
\beq
\bar s_n = \frac{\int \rho_n(x)^2 s(x)d^4x}{\int \rho_n(x)^2d^4x}, \qquad \rho_n(x) = |\psi_n(x)|^2
\eeq
that enhances the relevance of localised peaks. 
The data for some ensembles are reported in Figure~\ref{fig:ActionDensityNorm}, 
where we plot the ratio $\bar s_n/s$ ($s$ is the volume average of $s(x)$). 
To simplify the plots we averaged the results for modes in bins $[\lambda, \lambda+\delta]$. 
The lowest modes at all temperatures are localised with the peaks in regions with higher action. 
The average action ratio increases with the temperature. 
We observe again the effect of the boundary condition on the localisation. The average action as seen 
by the lowest modes with periodic boundary condition is suppressed (empty symbol), almost similar to the low temperature case. 

\begin{figure}[htp]
  \centering
  \includegraphics[width=.9\columnwidth]{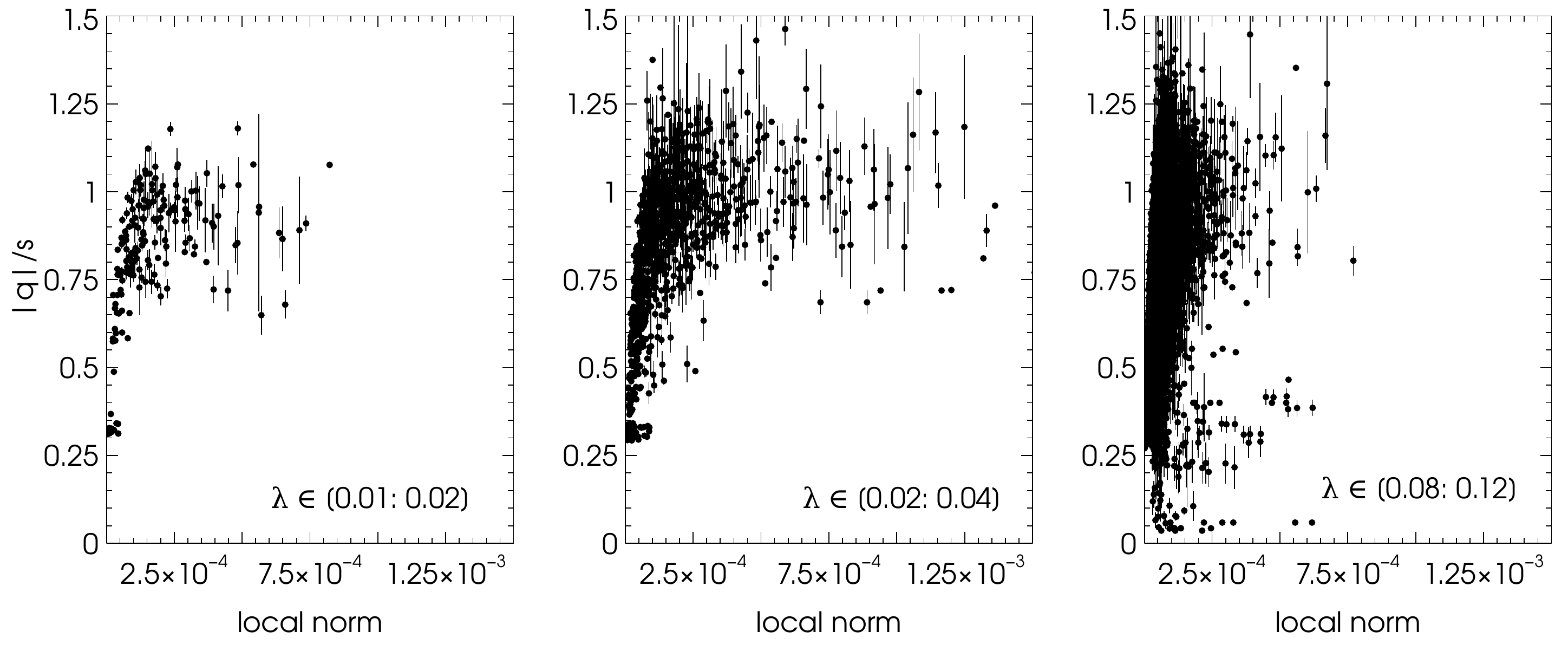}
  \caption{Dependence of the $|q|/s$ ratio on the norm for several eigenvalue sectors. Ensemble $32^3\times 12$, $\beta=4.30$, $m=0.01$, $T=220$ MeV. }
  \label{fig:ActionQNorm}
\end{figure}

A systematic view of the dominance of the regions where $|q|/s \sim 1$ is shown in Figure~\ref{fig:ActionQNorm}. 
For every eigenmode we averaged the quantity $|q|/s$ in a norm bin. The three panels represent different regions of the spectrum,
from the localised to the delocalised region. 

The picture that arises is that lowest modes are localised where the Polyakov loop is negative and precisely in the region where 
only two of its eigenvalues are equal and negative $\sim 1$, as discussed in Section~\ref{sec:PLBackgroundGauge}. The same regions are characterised by an action several times higher 
than the average and an almost self dual gauge field. These are some gauge invariant properties of the underlying fluctuations
supporting the low modes at finite temperature. On the same configurations the lowest modes of the operator with periodic boundary
condition show a much milder dependence, if none, on these features. The localisation mechanism depends on the boundary conditions. 

\subsection{Left and right projections}\label{sec:LeftRight}
In this section we present the study on the correlation between the localisation and the eigenmodes' chiral properties. The intent
is to find signals of a connection between the Anderson localisation mechanism and the chiral phase transition (via the Banks-Casher relation).

The non-zero eigenmodes of the non-hermitian Dirac operator $D$ have zero total integrated chirality:
\beq
\chi_n = \int \psi_n(x)^\dagger \gamma_5 \psi_n(x) d^4x = 0 
\eeq
thus any positive fluctuation of the local chirality $\psi_n(x)^\dagger \gamma_5 \psi_n(x)$ is balanced by a negative one. 
We identified the positive and negative fluctuations by projecting the eigenmodes onto their left- and right-handed components. 
The idea behind this separation is to study the chiral properties of the localised modes and the 
interaction between the left and right projections. 

\begin{figure}[ht]
  \centering
  \includegraphics[width=.8\columnwidth]{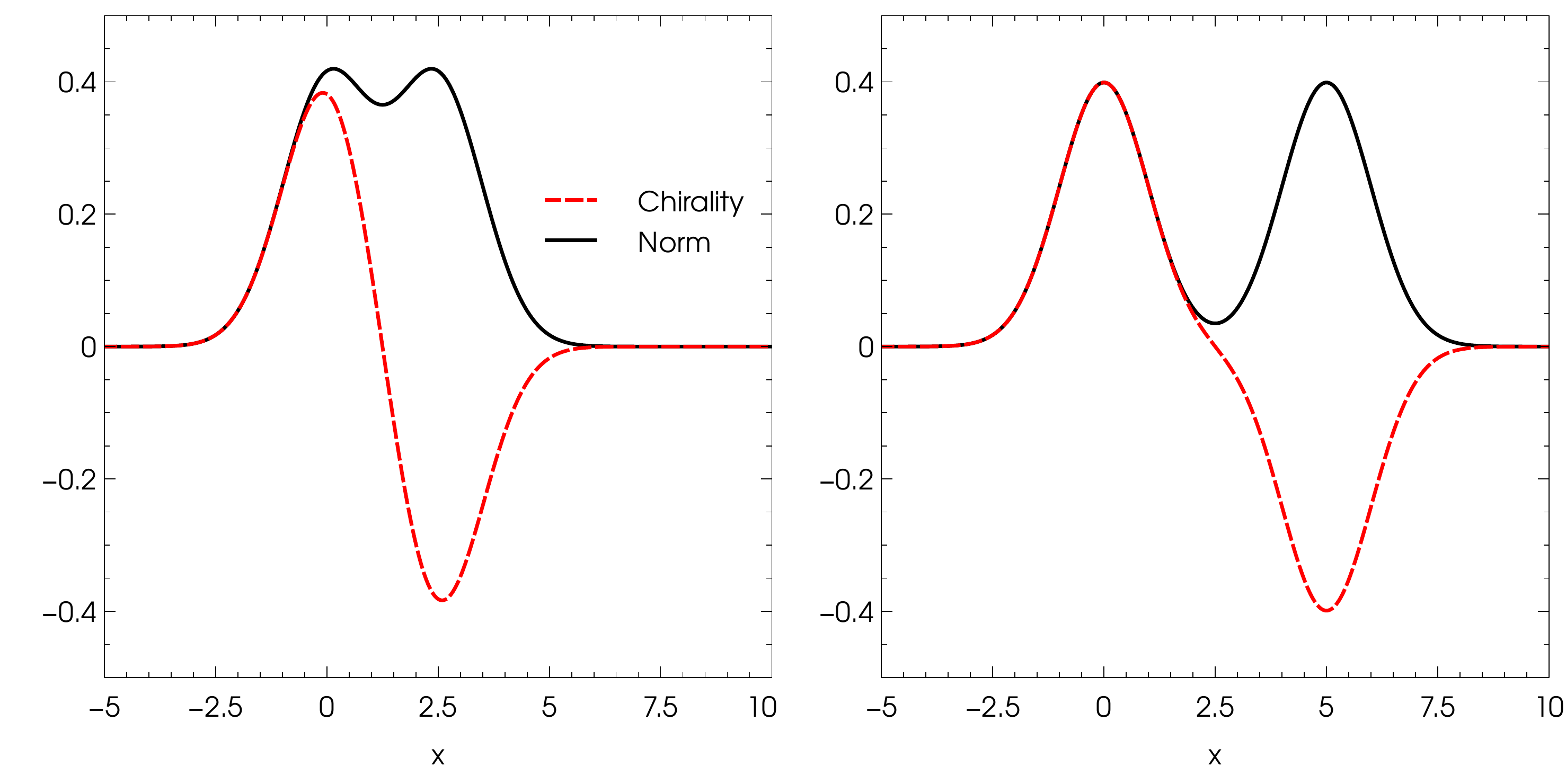}
  \caption{Simplified model description of a localised mode. Two gaussian distributions of the same weight and width represent the left and right components. 
The norm and the chirality are respectively the sum and the difference of the two gaussians. 
Left and right panels illustrate two cases depending on the separations.}
  \label{fig:Gaussians}
\end{figure}

For localised modes the typically observed spatial structure is a single large fluctuation of the norm density $\rho(x)$ that contains two
internal fluctuations of the chirality. To clarify this point, the ideal situation is depicted in Figure~\ref{fig:Gaussians} as a sum and difference 
of two gaussians at different separations.
Notice that this ideal situation is never perfectly realised, since several peaks can be present in the eigenmode landscape. 
Nevertheless it is an helpful simplified model that drives this analysis.  

The interaction between the left and the right projections can be parametrised by the spatial overlap among the two distributions. 
Defining $\rho_{L,R}$ as the square norm of the left and right projection of every eigenmode of $D$,
\beq
\psi_{L,R}(x) = \frac{1\pm \gamma_5}{2} \psi(x), \qquad \rho_{L,R} = |\psi_{L,R}(x)|^2,
\eeq
we construct the relative overlap as follows
\beq\label{eq:Overlap}
O_n = \frac{ \int \rho_{n,L}(x)\rho_{n,R}(x)d^4x}{ \frac{1}{2}\Bigl[ \int (\rho_{n,L}(x)^2 + \rho_{n,R}(x)^2) d^4x \Bigr]}, \qquad O_n \in [0,1],
\eeq
that represents the integrated probability of finding both projections at a given site normalized by their average size, their participation ratio. 
This is zero for a zero mode by definition.

When two infinitely distant zero modes, with opposite topological charge, are brought closer
the interaction of their wave functions will raise their eigenvalues from zero.  It is a simple system 
described by a $2\times 2$ matrix with only off-diagonal matrix elements, the modes interaction.
We parametrised the mode interaction with the $O_n$.
In this simplified picture the left and right modes increase their eigenvalue with the overlap. 

\begin{figure}[ht]
  \centering
  \includegraphics[width=.48\columnwidth]{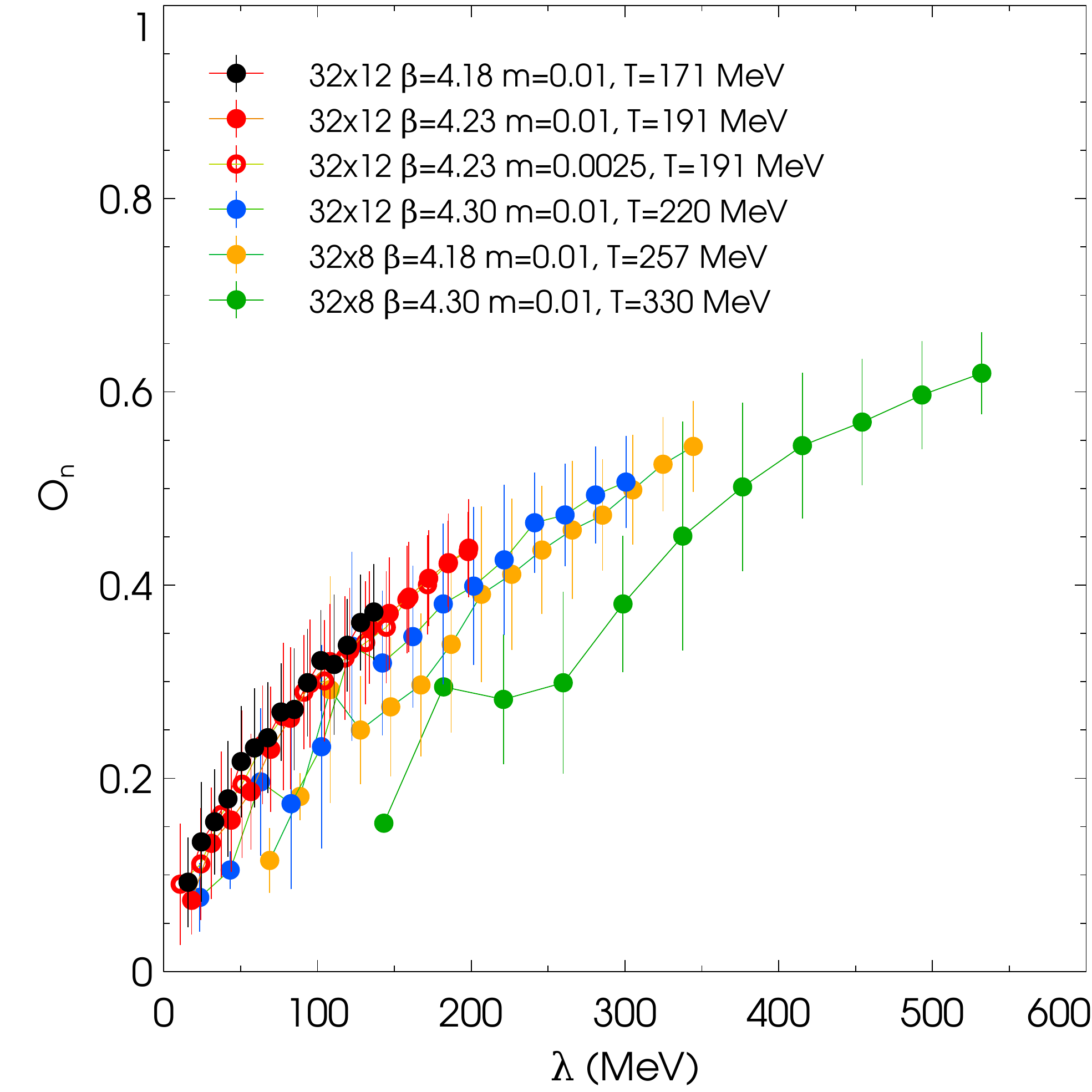}
  \includegraphics[width=.48\columnwidth]{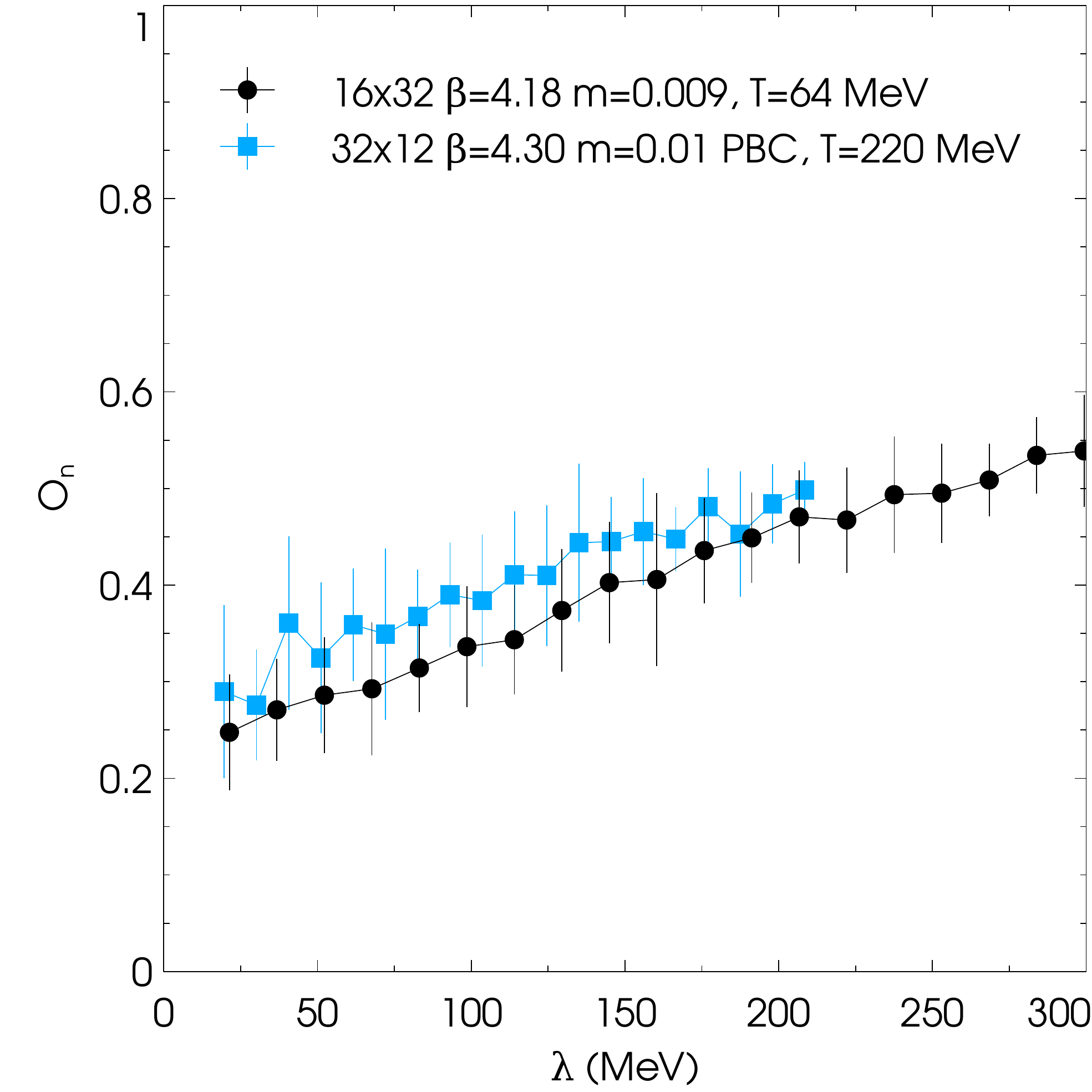}
  \caption{Overlap of the left and right projection for several ensembles, eq.~(\ref{eq:Overlap}). 
    Left panel shows the result for ensembles with localised modes. Right panel for ensembles with only delocalised modes.}
  \label{fig:Overlap}
\end{figure}

The numerical results for the high temperature ensembles are reported in Figure~\ref{fig:Overlap}. 
We plot the dependence of $O_n$ on the eigenvalue in the physical scale, after averaging
in eigenvalue bins. The most relevant feature is the monotonic increase of the 
overlap, with respect to the eigenvalue. 
There is a mild dependence on the temperature in a wide range $[0.9,1.9]T_d$ and a negligible dependence on the quark mass. 

In the second panel, we show the same quantity for ensembles at zero temperature, and for the operator 
with periodic boundary conditions on finite temperature configurations. 
Non-zero overlap is observed ever for near-zero modes, and this is not surprising 
due to the delocalised nature of the eigenmodes for these ensembles. 
However the monotonic behaviour persists. 

\begin{figure}[ht]
  \centering
  \includegraphics[width=.9\columnwidth]{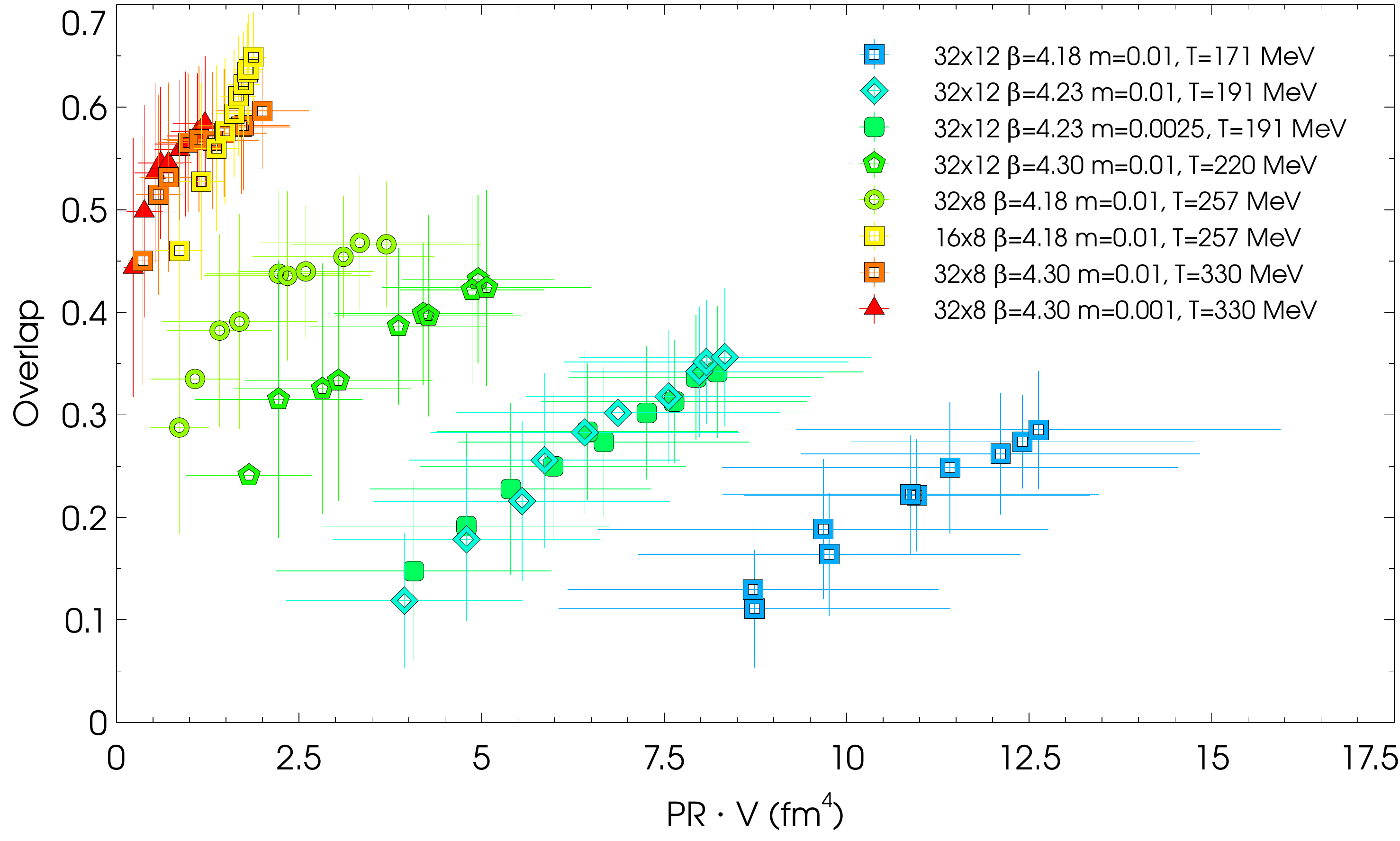}
  \caption{Dependence of the overlap $O_n$ of the left and right components on the physical size of the localised modes. The averages
of the first 10 non-zero modes are plotted.}
  \label{fig:PR_Overlaps}
\end{figure}

To accumulate further knowledge on the relation between the localisation and the left-right mode separation we considered the dependence 
of the overlap $O_n$ on the participation ratio. The $PR_n$ has been rescaled
by the volume $V$ to get the effective volume occupied by an eigenmode. In order to compare different ensembles we computed 
$V\cdot PR_n$ in physical units. The plot in Figure~\ref{fig:PR_Overlaps} shows the results
for ensembles where the lowest modes are localised. Only in these ensembles the rescaled PR 
has a physical meaning as the size of the localised mode (i.e. independent of the total volume, see Figure~\ref{fig:PR_Lowmodes}).

For every ensemble we plotted the averages of $V\cdot PR_n$ and the overlap $O_n$ for the first 10 non-zero eigenmodes. 
The ensembles are ordered by increasing temperature. There is an evident correlation between the size of 
the eigenmode and the overlap of its left and right components. As the temperature increases, the 
typical size shrinks while the overlap increases. 
This observation, in combination with the monotonic relation between the overlap and the eigenvalue,
pushes the eigenvalues up in the spectrum. 
The conclusion of this section is that the increasing localisation of the low modes is responsible for the diminishing spectral density around the
origin in the chirally symmetric phase. 


\section{Interpretation in terms of topological objects\label{sec:BPSMonopoles}}

We now turn to an interpretation of the results presented in the previous sections. 
Anticipating in short our conclusions, the background configurations 
supporting the near-zero modes have the characteristics of monopole-instanton (sometime called dyon in the literature) pairs.
In order to keep the paper self-contained, we first summarise some basic properties 
of the topological objects called calorons, and their constituents, the monopole-instantons. 
For a detailed discussion on their construction 
and their relation to confinement we recommend some reviews,
 articles \cite{Diakonov:2009jq, Bruckmann:2003yq, Shuryak:2012aa, Poppitz:2012nz} and the references therein. 

At zero temperature the relevant gauge field topological fluctuations are the instantons. Instantons are 
saddle-point configurations that satisfy the classical equation of motion  $D_\mu F_{\mu\nu} = 0$~\cite{Belavin:1975fg}. They also support zero modes of the 
Dirac operator~\cite{Atiyah:1963zz, Atiyah:1968mp}. 
It is straightforward to extend such solutions in the case of trivial holonomy by making the fields periodic~\cite{Harrington:1978ve, Harrington:1978ua}.
However, at finite temperature these are no longer solutions and they were extended 
by Kraan, van Baal and Lee, Lu~\cite{Kraan:1998pm, Kraan:1998kp, Kraan:1998sn, Lee:1998vu, Lee:1998bb, Lee:1997vp} to accommodate
a background with a non-trivial holonomy. Such saddle-point solutions are called calorons, KvBLL calorons from the authors. KvBLL proved that the 
caloron solution has monopole constituents with fractional topological charge~\cite{Kraan:1998sn}. 
These constituent monopoles are BPS monopoles and are static, time-independent
self-dual solutions of the equations of motion. 
They possess a non-zero chromomagnetic charge and correspond to non-trivial eigenvalues of the holonomy.

A caloron with unit topological charge in $SU(N)$ is a magnetically neutral combination of $N$ monopoles.
There are $N-1$ BPS monopoles, which we call $M$-type in this section, each carrying unitary magnetic charge in one of the Abelian Cartan subgroups, 
associated with the simple roots of the $SU(N)$ group (forming the $A_{N-1}$ Lie algebra):
\begin{equation}
A_{N-1} = \{\alpha_i | \alpha_i = e_i - e_{i+1}\}, \qquad i \in \{1 \dots N-1\}
\end{equation}
in which $e_i$ is the unit vector in $\mathbb R^N$ along the direction $i$. 
The $N$-th monopole has magnetic charge in each of these subgroups and it is associated with the affine root
\begin{equation}
\alpha_N = e_N - e_1 = - \sum_i^{N-1} \alpha_i.
\end{equation}
The latter is the so called Kaluza-Klein (KK) monopole~\cite{Lee:1997vp}, which we call $L$-type, and has a twisted gauge field along the compact dimension, 
so it is not time-independent. 
To fix the notation let us write the expressions for the asymptotic holonomy phases, $\mu_m \in [-\frac{1}{2}, \frac{1}{2}]$
\begin{eqnarray}
L(\infty) = \exp \Bigl [2\pi i\,\diag(\mu_1, \mu_2, \dots, \mu_N) \Bigr ],\nonumber\\
\mu_1 \leq \mu_2 \leq \dots \leq \mu_N \leq \mu_1 +1 \equiv \mu_{N+1}, \qquad \sum_{m=1}^N \mu_m = 0.
\end{eqnarray}
after a suitable gauge transformation to order the eigenvalues.
A summary of the properties of the monopole solutions is given in Table~\ref{table:BPS}, where $\nu_m$ 
is the difference between two consecutive eigenvalues of the holonomy, $\mu_m$ and $\mu_{m+1}$,
\beq
\nu_m \equiv \mu_{m+1} - \mu_{m}, \qquad \sum_{m=1}^N \nu_m = 1,
\eeq
the last relation coming from the determinant constraint. The naming convention for the monopoles adheres to~\cite{Diakonov:2009jq}.  
 The value of the gauge field at the centre of a $m$ monopole has been 
derived for all simple Lie groups by~\cite{Weinberg:1979zt, Davies:2000nw} and for $SU(N)$ is:
\beq
L(0) = \exp \Bigl [ 2\pi i\,\diag\Bigl (\mu_1, \mu_2, \dots, \mu_{m-1}, \underbrace{\frac{\mu_m + \mu_{m+1}}{2}}_m, \underbrace{\frac{\mu_m + \mu_{m+1}}{2}}_{m+1}, \dots \Bigr) \Bigr],
\label{eq:PL_center}
\eeq
which is cyclic for the $N$-th KK monopole, i.e. $m=N$ and $m+1=N+1 \equiv 1$. At least two of the eigenvalues must coincide in the centre of a monopole. 

\begin{table}
\begin{center}
\begin{tabular}{|c|c|c|c|c|}
\hline
 & $M_m$ & $\bar M_m$ & $L$ & $\bar L$ \\
\hline
E-charge & + & + & $-(N-1)$ & $-(N-1)$ \\
B-charge & + & $-$ & $-(N-1)$ & $+(N-1)$ \\
action S, $\frac{8\pi^2}{g^2}$ & $\nu_m$ & $\nu_m$ & $1-\sum\nu_m$ & $1-\sum\nu_m$\\
top. charge Q  & $\nu_m$ & $-\nu_m$ & $1-\sum\nu_m$ & $\sum\nu_m-1$\\
\hline
\end{tabular}
\caption{\label{table:BPS}List of characteristics of the self-dual solutions for $SU(N)$ groups. 
Here the index $m \in [1 \dots N-1]$, distinguishes the BPS ($M$-type) monopoles. $\nu_m$ phase is in unit of $2\pi$. The bar denotes the anti-monopoles solutions. $L$-type monopoles are
the Kaluza-Klein monopoles. A $Q=1$ caloron is composed by the sum of the different $N-1$ $M$-type monopoles and the $L$-type. 
It is neutral and has the canonical action $\frac{8\pi^2}{g^2}$. $L$ monopoles have B-charge $\pm 1$ in each one of the $N-1$ Cartan subgroups. 
 }
\end{center}
\end{table}

Below the phase transition the average Polyakov line is zero and thus the expected values for $\nu_m$ are all equal to $1/N$ on average, 
the maximal distance between all the eigenvalues. In this symmetric case the  
action and the topological charge are the same for all types of monopoles. Above the phase transition the Polyakov line tends to a 
configuration where all the eigenvalues are identical to one of the elements of the centre of $SU(N)$. This results in $M$-type monopoles to be light, $\nu_m \ll 1$, and $L$-type monopoles to be heavy, $\nu_N \sim 1$, see Table~\ref{table:BPS}.
The monopoles are not particles, therefore do not have a mass. We keep using the wording light/heavy with the caveat that it only refers to their carried action.

The topological charge fluctuation will be mostly localised around the heavy monopoles. 
A caloron is composed by the sum of the different $N-1$ $M$-type monopoles and one of the $L$-type. 
It is neutral, has the canonical action $\frac{8\pi^2}{g^2}$ and carries topological charge 1.

\begin{figure}
  \centering
  \includegraphics[width=.5\textwidth]{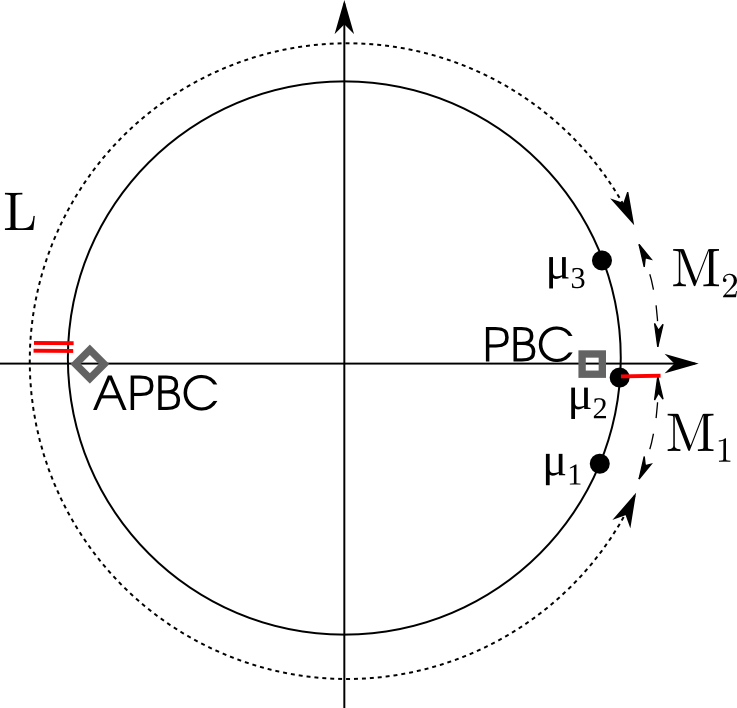}
  \caption{Example of holonomy eigenvalue $\mu_m$ (black dots) in the unit circle for $T\rightarrow \infty$. The dashed lines show the difference $\nu_m$ that are related to the action and the topological charge. The grey square and diamond indicate the fermionic phase for the respective boundary conditions. In this specific example the zero mode for APBC will be sitting in the $L$ monopole and on the $M_2$ monopole for PBC. The value of the holonomy at the centre of a specific monopole is read by taking the middle point of the corresponding arc (doubly degenerate) and the remaining eigenvalue (red lines in the figure showing the case of the $L$ monopole). }
  \label{fig:MonopolesDrawing}
\end{figure}

For the sake of clarity let us write the explicit form of the above equations for the case of an $SU(3)$ gauge theory by referring to Figure~\ref{fig:MonopolesDrawing}. The holonomy at spatial infinity at very large temperature approaches
\beq
L(r\rightarrow \infty)_{T\rightarrow \infty} = \exp \Bigl [2\pi i\,\diag(0,0,0) \Bigr ]
\eeq 
where $\mu_1 = \mu_2 =\mu_3 = 0$, and $r$ is the distance from the monopole core. The corresponding differences are $\nu_1 = \nu_2 = 0$ that give the 2 $M$-type monopoles, and $\nu_3 = 1$ for the heavy $L$ monopole. 
The action is zero for the two light monopoles and it is concentrated on the heavy one. Also the topological charge is carried only by the $L$ monopole. 
In this regime the holonomy at the centre of the monopoles, according to \ref{eq:PL_center} is given by
\beq
L(0)_M = \exp \Bigl [2\pi i\,\diag(0,0,0) \Bigr ], \qquad 
L(0)_L = \exp \Bigl [2\pi i\,\diag \Bigl (\frac{1}{2},\frac{1}{2},0 \Bigr ) \Bigr ]
\eeq 
modulo gauge rotation of the eigenvalues. They give $P=(1,0)$ and $P=(-1/3,0)$ respectively for the complex Polyakov line. 
This is of course an idealised situation, it is nevertheless a useful guide to understand the numerical data.  

The relation between the calorons and the zero modes of the Dirac operator 
has been explored analytically in~\cite{GarciaPerez:1999ux, Chernodub:1999wg, Nye:2000eg, Poppitz:2008hr}, where the exact expression 
for $SU(2)$ and $SU(N>2)$ fermionic zero-modes was derived for a caloron with $Q=1$ and non-trivial holonomy (see also~\cite{GarciaPerez:2009mg} for a discussion on the 
modes of the Dirac operator in the adjoint representation). 
A remarkable result of these papers is the dependence of the eigenmode location from the boundary conditions of the fermionic field. 

\begin{figure}
  \centering
  \includegraphics[width=.49\textwidth]{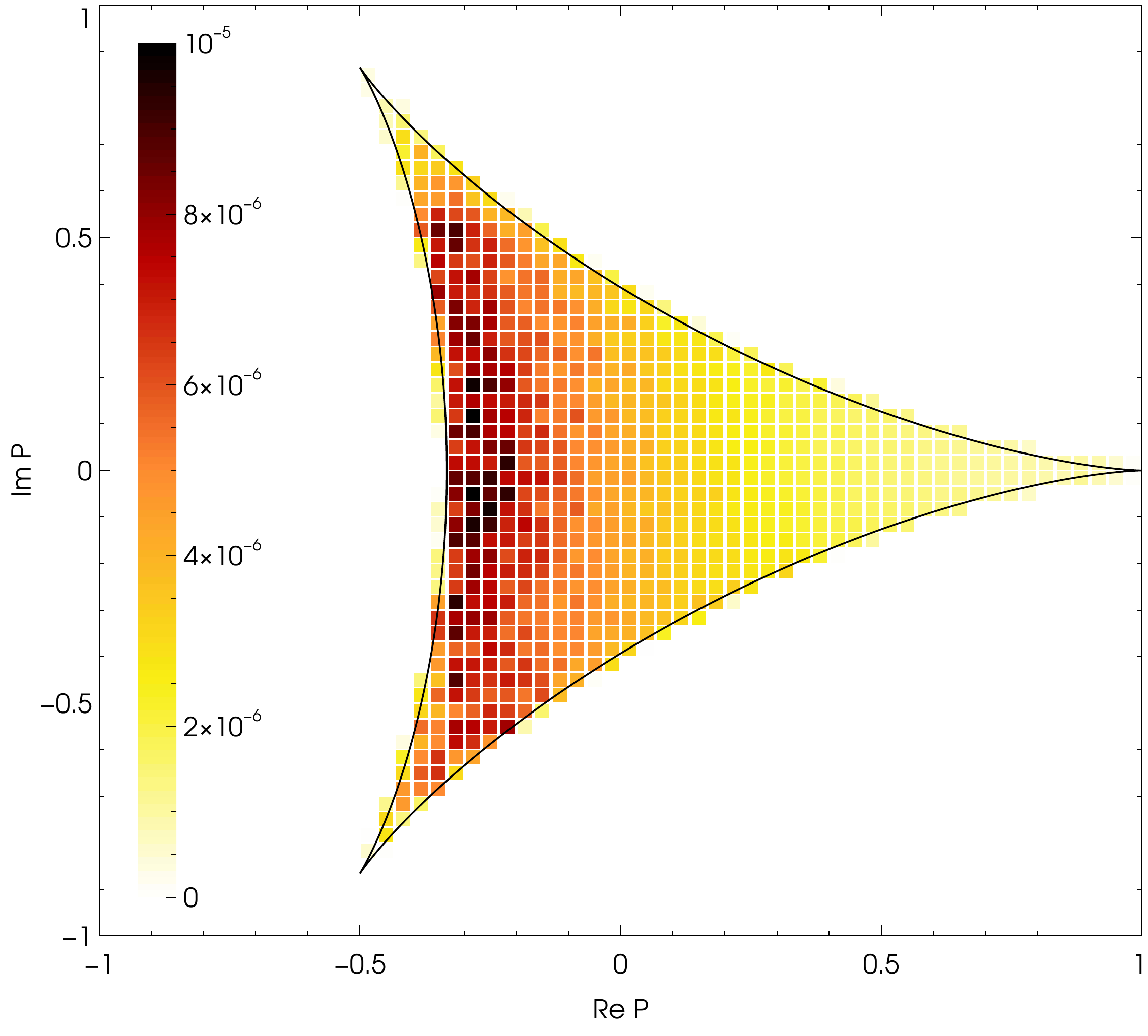}
  \includegraphics[width=.49\textwidth]{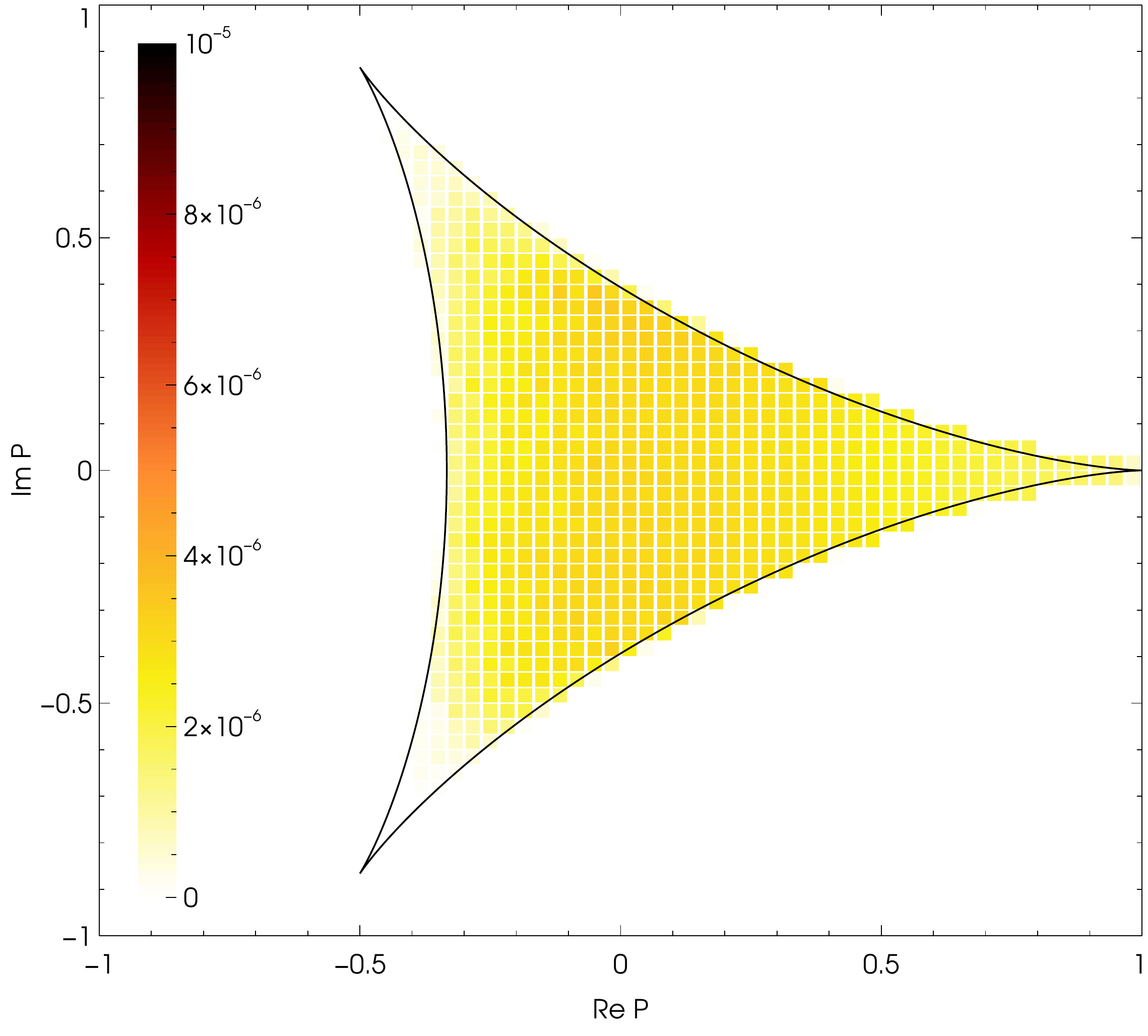}
  \caption{Density plot for the norm of the eigenmodes in the Polyakov loop plane. Average of the mode by mode results of Figure~\ref{fig:PLcorr}. Ensemble $32^3\times 12$, $\beta=4.30$, $m=0.01$, $T=220$ MeV. (left) Lower eigenmodes with $\lambda < 0.03$. (right) higher eigenmodes with $\lambda > 0.1$. Analogous results by filtering using the eigenmodes in~\cite{Bornyakov:2014esa, Bornyakov:2015xao}.}
  \label{fig:PLdistribution}
\end{figure}

The situation is best illustrated in the case of $SU(2)$ where 
only two species form the caloron~\cite{GarciaPerez:1999ux, Poppitz:2008hr}. 
Let's start with a background configuration for a caloron made of the combination~$ML$.
The zero mode is localised on one of the constituents and 
distinguishes between the periodic and anti-periodic boundary conditions. 
The general result states that for the boundary condition $\psi(t+T) = \exp(2\pi i \phi) \psi(t)$
the eigenmode is localised on the monopole $m$ that satisfies $\phi \in [\mu_m, \mu_{m+1}]$.
At high temperature periodic zero modes are localised on the light $M$-constituents 
while anti-periodic zero-modes localise around the heavier $L$ monopoles, see Figure~\ref{fig:MonopolesDrawing}. 
This is true for $SU(N)$ groups too, with some exceptions~\cite{Chernodub:1999wg} that are irrelevant in the deconfined phase. 
Figure~\ref{fig:MonopolesDrawing} exemplifies the case of $N_c=3$ and
Figure~\ref{fig:PLdistribution} depicts the typical distribution of the 
norm of the eigenmodes in the Polyakov loop plane 
for the low-mode and the high-mode regions in the high temperature phase.

The results of this paper suggest that these properties of the zero-modes for the classical Yang-Mills (YM) background do apply also for 
low-lying non-zero modes. The underlying assumption is that topological fluctuations are stable under deformations. 
The numerical evidences in Section~\ref{sec:Analysis}, on the properties of the background configurations supporting the low-lying modes
with anti-periodic boundary conditions, indicate
that these modes are localised around the heavy KK monopole fluctuations of the gauge field.
We measured a set of gauge invariant observables characterising the gauge fluctuations. The values 
of the Polyakov line at the centre of the eigenmodes, the averaged action, the self-duality of the gauge field and the boundary 
condition dependence are all in accordance with the characteristic of the monopole-instantons. 
We also cross-checked the dependence on 
the boundary condition by confirming that the number of zero modes is unchanged by the boundary condition. 

These monopole-instantons are not necessarily part of a caloron:  
in the case they support non-zero modes, the results from Section~\ref{sec:Analysis} 
support the picture that they are coupled in topologically neutral pairs, $L \bar L$ pairs.
Other works in the literature have presented evidence of monopole-instanton (dyonic) structures near the phase transition by using the 
low modes as UV filter for the topological charge, see \cite{Bornyakov:2014esa, Bornyakov:2015xao} and references therein.
Recently a work with chiral fermions investigated the topological features of the high temperature phase~\cite{Dick:2015twa}. 

Here we showed using chiral fermions that, mode by mode, the disorder underlying the Anderson localisation 
is composed by $L$-type monopoles which have increasing action
with the temperature, and this is reflected by the increasing localisation of the near-zero modes. 
At finite temperature these heavy $L$-type monopoles excitations are expected to be suppressed 
while most of the lattice points support lighter fluctuations.
This is confirmed qualitatively by our results, but we did not estimate the $L$ monopole density.
The situation changes below the phase transition where there is no difference in the monopole actions on average, and the 
$L$-type monopoles are now easier to appear as fluctuations.
If the low-modes support mechanism is the same on both sides, it is not difficult to qualitatively 
explain the delocalisation below $T_d$, and above $T_d$ by
changing the boundary conditions, in connection with the expected relevant monopole abundance.

In this paper we do not address the task of estimating the abundance of the monopole structures and its dependence on the 
temperature and mass of the sea quarks. This will be the subject of a future study. The statistical mechanics 
of monopole instanton ensembles has been studied in~\cite{Faccioli:2013ja}.

The relations of other pairs of monopoles ($L \bar M$, $M \bar L$, $M \bar M$) to the lowest eigenmodes is
 disfavoured by the numerical evidence (with APBC). The support of the 
near-zero modes strongly favours large action for both positive and negative topology fluctuations. It also favours a 
region where the Polyakov loop has only two coinciding eigenvalues (${\rm Re}P\rightarrow-1/3$). All the other couples except $L\bar L$ do
not satisfy these requirements. This does not exclude their presence in the gauge configurations 
but it rules out their relation to the Anderson localisation mechanism.
By changing the boundary conditions the situation is reversed and the eigenmodes are allowed to hop
between $M$-type monopoles. Results still seem to exclude that mixed pairs are correlated to the low energy spectrum. 

An interesting result is that we could not observe a strong dependence of the results on the fermion masses. 
If there is any dependence, it is below the sensitivity of our probes. 
A mass dependence of the interaction among monopole pairs was predicted in some models~\cite{Shuryak:2012aa}.\footnote{For discussion of the dyon-antidyon liquid model 
see~\cite{Liu:2015ufa, Liu:2015jsa}.}

We have observed in Section~\ref{sec:LeftRight} that the increasing localisation with the temperature, coming from the heavy monopoles, 
is associated with an increasing overlap
of the lowest modes. 
In the monopole picture two infinitely distant $L$-type monopoles will support two zero modes, when surrounded by $M$-type monopoles.
Any overlap of their wave function will result in an increased eigenvalue. 
This picture explains the measured monotonic dependence between the eigenvalue  $\lambda_n$ and the overlap $O_n$.
Together with the localisation-overlap relation, this can be the mechanism that 
suppresses the spectral density at high temperature, eventually leading to the chirally symmetric phase in the thermodynamical limit.
This is a conjecture that needs further investigation. If proven true can be 
the leading indication of a deeper connection between the deconfinement phase transition and the chiral phase transition, via localisation and Banks-Casher relation. 
In principle two distinct mechanism are at work here that can drive the chiral restoration. The first is the abundance of $L$-monopoles
that is suppressed with the temperature. The second one is the formation of $L\bar L$ molecules that can lead to a spectral gap, Figure~\ref{fig:Overlap}.
The current numerical results are not sufficient to differentiate among the two and clarify their relative importance.

Below the phase transition the action of the monopoles is similar and thus the relative abundance of $M$ and $L$ monopoles would be the same. The consequence is an
independence of the results on the boundary conditions. The opposite case happens at finite temperature, as demonstrated by the data in Section~\ref{sec:Analysis}. 
Models have been discussed in~\cite{Shuryak:2012aa} that reached similar conclusions. The dependence of the chiral condensate on the 
boundary conditions in the deconfined phase has been reported in~\cite{Bilgici:2008qy, Bilgici:2009tx}.
In this conjecture the change in density of the $L$ monopoles, strictly intertwined with the Polyakov line, is the trigger of the chiral phase transition and 
can explain our results of Section~\ref{sec:LeftRight} as well as~\cite{Bilgici:2008qy, Bilgici:2009tx}, hence $T_c \geq T_d$.
Notice that the Anderson localisation and the presence of monopoles~\cite{Bornyakov:2014esa} 
has been observed also in pure gauge setups~\cite{Giordano:2015vla} where all fermion interactions are switched off. This seems to suggest that the fermion interactions 
could be a second order effect in the mechanism of localisation.

\section{Conclusions and future works\label{sec:Conclusions}}

We have studied the finite temperature phase of two-flavor QCD using chiral fermions around the phase transition, $T \in [0.9,1.9]T_c$.
At fixed finite temperature $T$, the Dirac spectrum undergoes a quantum phase transition that is in accordance with the prediction of the 
Anderson model for localisation~\cite{Giordano:2013taa, Ujfalusi:2015nha}. 
The near-zero Dirac eigenmodes at high temperature are localised up to a critical eigenvalue $\lambda_c(T)$.
The Anderson model for the metal-insulator transition 
describes a localisation mechanism 
of wave functions due to the presence of disorder in the crystal. 
We studied the properties of the gauge configurations that can
cause disorder and localise the lowest Dirac eigenmodes.

We presented several numerical evidences that the background gauge configurations supporting each near-zero eigenmode of 
the Dirac operator in the fundamental representation are self-dual and carry higher action than 
the volume average. 
They are strongly related to fluctuations of the Polyakov line where two of the eigenvalues 
of the holonomy are identical (equal to $-1$ at high temperature).
The data show localisation of the lowest eigenmodes in such 
regions of negative Polyakov line, large action and self-dual gauge fields. 
Changing the boundary conditions of the Dirac operator on the same finite temperature ensembles
washes up the localised states and all modes become delocalised. 

Furthermore, we studied the chiral properties and in particular the overlap of the left and right chiral projections of the non-zero eigenmodes.
This overlap is a monotonically increasing function of the eigenvalue with mild dependence on the temperature. 
By varying the temperature the overlap shows also inverse correlation with the participation ratio, meaning that the more localised
modes show larger overlap of their left and right components. 

We proposed an interpretation of the numerical results in terms of monopole-instanton fluctuations of the 
background gauge field. Monopole-instantons are self-dual solutions of the Yang-Mills equations of motion
in a non-trivial holonomy background. In the deconfined phase there is a class of monopoles that 
support larger action (in our terminology ``heavier''). The properties of the observed
gauge background are in accordance with the properties of heavy monopole-antimonopole molecules. 
These fluctuations trap the lowest eigenmodes and trigger localisation. 

In the monopole model the support of the eigenmodes is expected to change with the boundary conditions. At high temperature, heavier monopoles
are favoured by the physical case of anti-periodic boundary conditions. The eigenmode
support moves to the lighter monopoles with periodic boundary conditions. 
We observed a dependence on the boundary conditions of the eigenmodes consistent with this expectation. 

Based on the current results, we conjectured a relation of the deconfinement phase transition, 
driven by the Polyakov line, and the chiral phase transition. 
The key connection element is the heavy monopole action that is linked to the localisation of the low-modes,
 and the relation of the overlap between the left and right components of the eigenmode and its eigenvalue.
Further study is needed to strengthen the connection and the 
next step would be a proper estimate of the density of the several monopole species 
and monopole-pairs and the dependence of these characteristics on the temperature and mass of fermions. 
The connection between the localisation mechanism and the chiral phase transition has been considered
in \cite{GarciaGarcia:2006gr} and in the model based approach \cite{Pittler:2014qea, Giordano:2016cjs}. 

We are currently investigating a non-perturbative derivation of an \emph{effective potential} 
that drives the localisation of the fermionic modes. 
This would be a particularly useful tool to understand Anderson localisation mechanism in QCD, 
especially the appearance of a mobility edge at finite temperature. 
The results will appear in future works. 

An interesting extension of this study would be the research on the adjoint fermions case. 
The two phase transition
are separated in the $SU(3)$ case~\cite{Karsch:1998qj}. The different localisation properties 
of the adjoint eigenmodes~\cite{GarciaPerez:2009mg} can 
account for this difference.
 
SU(3) with adjoint fermions in $R^3\times S^1$~\cite{Cossu:2009sq, Misumi:2014raa, Cossu:2013ora} 
is another theory that has a peculiar phase structure where the average Polyakov line can 
get non-trivial values. In the completely broken $U(1)\times U(1)$ phase where 
new magnetically charged and topologically neutral molecules of BPS and anti-KK monopoles, 
called bions are expected to
be relevant for confinement~\cite{Unsal:2007jx, Misumi:2014raa}. 
It would be interesting to test the background configurations for bions using the methods illustrated in this paper. 

\begin{acknowledgments}
We thank the members of the JLQCD collaboration, especially H. Fukaya 
and A. Tomiya, for many discussions and for sharing the work of 
generating and analyzing the finite temperature lattice ensembles.
GC would like to thank M. Giordano, M. \"Unsal and T. Kanazawa 
for several very useful discussions during the work on this subject 
and for helping to improve  the original manuscript. 

Numerical simulations are performed on IBM System Blue Gene Solution at 
High Energy Accelerator Research Organization (KEK) under a support for 
its Large Scale Simulation Program (No. 14/15-10). 
This work is supported in part by the Grant-in-Aid of the 
Ministry of Education (No. 26247043, 15K05065) and by MEXT SPIRE and JICFuS.
\end{acknowledgments}

\bibliography{references}

\end{document}